\documentclass[pre,twocolumn, aps]{revtex4}
\def\llsymbol#1{\@llsymbol{\@nameuse{c@#1}}}
\def\@llsymbol#1{\ifcase#1\or {}\or {'}\or {''}\or {'''}\or
   {''''}\or {'''''}\or  \else\@ctrerr\fi\relaz}
\newcounter{contador}
\newcommand{\letra}{
   \stepcounter{equation}
   \setcounter{contador}{\value{equation}}
   \setcounter{equation}{0}
   \renewcommand{\theequation}{\thecontador\alph{equation}}}
\newcommand{\antiletra}{
   \renewcommand{\theequation}{\arabic{equation}}
   \setcounter{equation}{\value{contador}}}

\usepackage{color}
\usepackage{indentfirst}
\usepackage{eucal}
\usepackage{amsmath}
\usepackage{amsfonts}
\usepackage{amssymb}
\usepackage{mathrsfs}
\begin{document}
\title{A limit of the confluent Heun equation and the Schr\"odinger equation
for an inverted potential and for an electric dipole}
\author{L\'ea Jaccoud El-Jaick}
\email{leajj@cbpf.br}
\author{Bartolomeu D. B. Figueiredo}
\email{barto@cbpf.br}
\affiliation{Centro Brasileiro de Pesquisas F\'{\i}sicas (CBPF),\\
Rua Dr. Xavier Sigaud, 150, CEP 22290-180, Rio de Janeiro, RJ, Brasil}
\begin{abstract}
\noindent
We reexamine and extend a
group of solutions in series of Bessel
functions for a limiting case of the confluent Heun
equation and, then, apply such solutions to the
one-dimensional
Schr\"odinger equation with an inverted
quasi-exactly solvable potential
as well as to the angular equation
for an electron in the field of
a point electric dipole. For the first problem we find finite- and
infinite-series solutions which are convergent and bounded for any value of the
independent variable. For the angular equation,
we also find expansions in series of Jacobi
polynomials.
\end{abstract}
\maketitle

\section{Introduction}
Firstly we revise and extend some solutions in series of Bessel
functions for a limiting case of the confluent
Heun equation (CHE) -- the CHE is also called generalized
spheroidal wave equation \cite{fisher,leaver1,wilson1}.
Then we apply these solutions to the one-dimensional
Schr\"odinger equation with an inverted quasi-exactly
solvable potential considered by Cho and Ho
\cite{cho,cho-2}, as well as to the angular equation
for an electron in the field of a point electric dipole \cite{levy}.

For the inverted potential we find
even and odd eigenstates
given by finite series of Bessel functions corresponding to
the same energy eigenvalues, for any value of
the parameter which characterizes the quasi-exact solvability.
We obtain as well infinite-series solutions which are convergent
and bounded over the entire range of the independent variable.

For the angular equation, in addition to expansions in
series of Bessel functions, we get solutions in series of
Jacobi polynomials,
equivalent to the ones proposed by Alhaidari \cite{alhaidari}.
The latter type of solutions is inapplicable
to the inverted potential and this is the reason for
regarding expansions in series of Bessel functions.

The  stationary one-dimensional Schr\"{o}dinger
equation for a particle with mass $M$
and energy $E$ is written as
\begin{eqnarray}
\label{schr}
\frac{d^2\psi}{du^2}+\big[{\cal E}-V(u)\big]\psi=0
\end{eqnarray}
with
\begin{eqnarray}
 u={a} x,
 \quad  {\cal E}=\frac{2M E}{\hbar^2 a^2},
 \quad   V(u)=\frac{2M V(x)}
{\hbar^2 a^2},\nonumber
\end{eqnarray}
where ${a}$ is a  constant with inverse-length dimension,
$\hbar$ is the Plank constant
divided by $2\pi$, $x$ is the spatial coordinate
and $V(x)$ is the potential.
For the case considered by Cho and Ho
\begin{eqnarray}\label{v1}
V(u)=-\frac{\mathrm{ b}^{2}}{4}\sinh^{2}u-
\frac{l^2-(1/4)}{\cosh^{2}u},\\
\left[u\in(-\infty, \infty),
\quad l=1,2,3,\cdots\right]\nonumber
\end{eqnarray}
%
%
where $\text{b}$ is a positive real constant.
This is a bottomless potential
in the sense that $V(u)\rightarrow{-}\infty$
when $u\rightarrow\pm \infty$. If $\text{b}^2<4(l+1)^2-1$,
it is an inverted double well; if
$\text{b}^2\geq4(l+1)^2-1$, it resembles the
potential of an inverted oscillator.
Then, we have the equation
\begin{equation}\label{cho-ho}
\frac{d^{2}\psi}{du^{2}}+
\left[{\cal E}+\frac{\text{b}^2}{4}\sinh^2u +\left(l^2-\frac{1}{4}
\right) \frac{1}{\cosh^2u}\right]\psi=0,
\end{equation}
which, for the previous values assigned to $l$,
is a quasi-exactly solvable equation.

A quantum mechanical problem is
quasi-exactly solvable (QES) if one part of its
energy spectrum and the respective eigenfunctions
can be computed explicitly by algebraic methods
\cite{turbiner,turbiner1,ushveridze1}. From another
viewpoint \cite{kalnins}, a problem is QES if admits solutions
given by finite series whose coefficients
necessarily satisfy three-term  or higher order
recurrence relations, and it
is exactly solvable if its solutions are given by
hypergeometric functions
(two-term recurrence relations). This is suitable
for problems involving Heun equations because in general
these present finite-series solutions  \cite{ronveaux};
they admit as well infinite-series solutions which could
afford the rest of the spectrum.

As to the second problem, the Schr\"odinger equation for an
electron in the field of a point electric
dipole is used for modeling the scattering of negative ions by
polar molecules \cite{levy}. When it is {separated}
in spherical coordinates ($r,\theta,\varphi$),
the $\theta$-dependence satisfies the equation
\begin{equation}\label{theta}
\left[ \frac{1}{\sin{\theta}}\frac{d}{d\theta}\left(
\sin{\theta}\frac{d}{d\theta}\right)+C-\beta\cos{\theta}-
\frac{m^2}{\sin^2\theta}\right] \Theta(\theta)=0,
\end{equation}
where $C$ is a separation constant to be determined, $\beta$
is the dipole moment parameter and  $m$
is the angular momentum conjugate to
$\varphi$.

In sections III and IV, respectively, equations (\ref{cho-ho})
and (\ref{theta}) are transformed into a confluent
Heun equation (CHE) having the form
\begin{eqnarray}
\label{incegswe}
z(z-z_{0})\frac{d^{2}U}{dz^{2}}+(B_{1}+B_{2}z)
\frac{dU}{dz}+\nonumber\\
\left[B_{3}+q(z-z_{0})\right]U=0,\quad (q\neq0)
\end{eqnarray}
where  $z_0$, $B_{i}$ and $q$ are constants
(if $q=0$, the equation degenerates
into a hypergeometric equation). This equation is obtained
by applying the so-called Whittaker-Ince limit \cite{lea,eu}
\begin{eqnarray}\label{ince}
\omega\rightarrow 0, \quad
\eta\rightarrow
\infty \quad \mbox{such that }\quad  \ 2\eta \omega =-q,
\end{eqnarray}
to the following form of the confluent Heun
equation  \cite{leaver1}
\begin{eqnarray}
\label{gswe}
&&z(z-z_{0})\frac{d^{2}U}{dz^{2}}+(B_{1}+B_{2}z)
\frac{dU}{dz}+\nonumber\\
&&\left[B_{3}-2\eta
\omega(z-z_{0})+\omega^{2}z(z-z_{0})\right]U=0,
\end{eqnarray}
where $z_0$, $B_{i}$, $\eta$ and $\omega$ are constants.
In both cases, $z=0$ and $z=z_{0}$ are regular singular
points having  indicial exponents  ($0,1+B_{1}/z_{0}$)
and ($0,1-B_{2}-B_{1}/z_{0}$), respectively.
However, equations (\ref{incegswe}) and (\ref{gswe}) differ
by the behavior of their solutions
at the irregular singularity $z=\infty$, namely  \cite{eu},
\begin{eqnarray}
&&U(z)\sim
e^{\pm 2i\sqrt{qz}}z^{(1/4)-(B_{2}/2)} \ \text{for Eq. (\ref{incegswe})},\nonumber
\end{eqnarray}
\begin{eqnarray}\label{thome2}
&&U(z)\sim e^{\pm i\omega z}z^{\mp i\eta-(B_{2}/2)}\ \text {for Eq. (\ref{gswe})}.
\quad\
\end{eqnarray}

We shall refer to equation (\ref{incegswe}) as Whittaker-Ince

limit of the CHE (\ref{gswe}). This equation also appears
in the separation of the
variables for the Laplace-Beltrami operator in an
Eguchi-Hanson space
\cite{mignemi,malmendier} and
for the Schr\"odinger equation for an ion in the field of a
electric quadrupole in two dimensions \cite{alhaidari2}.
In addition, it includes the Mathieu
equation as a particular case {and, when $z_0\rightarrow0 $,} it
leads to a double-confluent Heun equation which arises,
for example, in the
scattering of ions by polarizable targets  \cite{eu}.

In reference \cite{eu} the limit (\ref{ince})
is used to find solutions for equation
(\ref{incegswe}) from solutions of the CHE
(\ref{gswe}). However, those solutions are not
sufficient to handle the previous problems.
For this reason, from known solutions
in series of Hankel functions for
equation (\ref{incegswe}),  in section II we construct
solutions in series of Bessel functions of the first and second kind.
We find that the expansions in terms of the functions of the
first kind, under certain conditions, are convergent
and bounded for any value of the
independent variable. The other expansions in general converge
for $|z|>  |z_0| $ or $
|z-z_0|> |z_0|$, but under special conditions converge also at $|z|=|z_0|$
or $|z-z_0|=|z_0|$.

In section III, using the expansions in series of Bessel functions
of the first kind, we obtain even and odd eigenfunctions
for the Cho-Ho equation (\ref{cho-ho}), given
by finite series and by infinite series as well.
The three-term recurrence relations
for the coefficients of the finite series imply
that even and odd states are degenerate. A proof that
this degeneracy takes place for any value of the parameter
$l$ is possible because now we have a general expression
for the eingenstates. Expansions in series of the other
Bessel functions also lead to solutions bounded for any value of $ z$,
but deprived of the parity property.

In section IV we apply the solutions in series of Bessel functions
for the angular equation of the
point dipole. {For this problem,
we also get solutions in series of Jacobi or associated
Legendre polynomials and point out why} this type of
expansions does not work for the inverted potential.

{Conclusions are reported in section V, while
Appendix A contains some formulas
used throughout the paper and Appendix B
exhibits solutions which have been omitted in section II
but are used in sections III and IV.}

%
\section{Solutions of the Whittaker-Ince limit
of the CHE}

All the solutions we will use for the Whittaker-Ince limit of the
CHE  may be generated by applying the limits
(\ref{ince}) on solutions of the CHE (\ref{gswe}),
but may also be derived directly from (\ref{incegswe}).
In this section we deal with a group of solutions given by
expansions in series of four Bessel functions which are denoted
by $Z_{\alpha}^{(j)}(x)$ ($j=1,2,3,4$),
according as \cite{arscott}
\begin{equation}
Z_{\alpha}^{(1)}(x)=J_{\alpha}(x),\ \
Z_{\alpha}^{(2)}(x)=Y_{\alpha}(x),\nonumber\qquad
\end{equation}
\begin{equation}\label{Z2}
Z_{\alpha}^{(3)}(x)=H_{\alpha}^{(1)}(x),\ \
Z_{\alpha}^{(4)}(x)=H_{\alpha}^{(2)}(x),
\end{equation}
where $J_{\alpha}$ and $Y_{\alpha}$
are the Bessel functions
of the first and second kind, respectively, whereas
$H_{\alpha}^{(1)}$ and $H_{\alpha}^{(2)}$
are the first and the second Hankel functions ($\alpha$
is called the order of the functions). Thus, a first set of
solutions has the form
\begin{eqnarray}\label{Z}
U_{1}^{(j)}(z) =z^{(1-B_{2})/2}
\displaystyle \sum_{n=0}^{\infty}(-1)^{n} b_{n}^{(1)}
Z_{2n+B_{2}-1}^{(j)}\left(2\sqrt{qz}\right),
\\
\quad [B_{2}\neq 0,-1,-2,\cdots].\nonumber
\end{eqnarray}
where the presence of the same coefficients $b_{n}^{(1)}$
is due to the fact that the four Bessel functions
satisfy the same difference and differential equations \cite{luke}.
These are called one-sided series because the summation
runs from zero to positive infinity.

In a region where the four expansions are valid, only
two of them are independent. For instance,
from equations (\ref{J-H}) and (\ref{Y-H}) we
get
\begin{equation}
U_{1}^{(1)}(z)=\frac{1}{2}
\left[ U_{1}^{(3)}(z)+U_{1}^{(4)}(z)\right],\nonumber
\end{equation}
\begin{equation}\label{combinacao}
U_{1}^{(2)}(z)=\frac{1}{2i}
\left[ U_{1}^{(3)}(z)-U_{1}^{(4)}(z)\right]
\end{equation}
up to a multiplicative constant. However, each
expansion presents different behaviors at $z=\infty$
as inferred from equation (\ref{Bessel-assimptotico}).

In section II.A we explain how the above
set of solutions has been obtained, write the general forms of
the three-term recurrence relations for the series coefficients
and present the transformations used to generate additional
sets of solutions. In section II.B we analyse some
properties of the solutions and discuss the case
of the Mathieu equation, while in section II.C we study the
convergence of the solutions.

\subsection{Recurrence relations and transformations of the equation}
The expansions in series of Hankel functions which appear
in (\ref{Z}) are taken
from Ref. \cite{eu} where they have been expressed
by series of the modified Bessel functions
$K_{2n+B_{2}-1}(\pm2i\sqrt{qz})$
as
\begin{eqnarray*} U_{1}^{\infty}(z) =z^{(1-B_{2})/2}
\displaystyle \sum_{n=0}^{\infty} b_{n}^{(1)}
K_{2n+B_{2}-1}\left(\pm2i\sqrt{qz}\right),\nonumber\\
\left[ B_{2}\neq 0,-1,-2,\cdots\right] .
\end{eqnarray*}
{In terms of Hankel functions \cite{erdelyi1b},}
\[\  K_{2n+B_2-1}(-2i\sqrt{qz})\propto (-1)^nH_{2n+B_2-1}^{(1)}(2\sqrt{qz}),\]
\[K_{2n+B_2-1}(2i\sqrt{qz})\propto (-1)^nH_{2n+B_2-1}^{(2)}
(2\sqrt{qz}),\ \]
where the proportionality constants do not depend on
the summation index $n$. Thus, the preceding solution
acquires the form of the solutions $U_{1}^{(3)}$
and $U_{1}^{(4)}$. The new solutions $U_{1}^{(1)}$
and $U_{1}^{(2)}$ result from the properties of the
Bessel functions, as aforementioned, and are
used in sections III and IV.

The recurrence relations for
series coefficients $b_n$ and the respective characteristic
equations (in terms of continued fractions) assume one of
the following forms \cite{eu}
%
\begin{eqnarray}\label{r1a}
\left.
\begin{array}{l}
\alpha_{0}b_{1}+\beta_{0}b_{0}=0,
\vspace{.2cm} \\
\alpha_{n}b_{n+1}+\beta_{n}b_{n}+
\gamma_{n}b_{n-1}=0\ (n\geq1)
\end{array}\right\} \Rightarrow\nonumber\vspace{.2cm}\\
\beta_{0}=\frac{\alpha_{0}\gamma_{1}}{\beta_{1}-}\ \frac{\alpha_{1}
\gamma_{2}}
{\beta_{2}-}\ \frac{\alpha_{2}\gamma_{3}}{\beta_{3}-}\cdots ,
\end{eqnarray}
\begin{eqnarray}
\left.
\begin{array}{l}
\alpha_{0}b_{1}+\beta_{0}b_{0}=0,
\vspace{.2cm} \\
\alpha_{1}b_{2}+\beta_{1}b_{1}+\left[
\alpha_{-1}+\gamma_{1}\right]b_{0}=0,
\vspace{.2cm} \\
\alpha_{n}b_{n+1}+\beta_{n}b_{n}+\gamma_{n}
b_{n-1}=0\ (n\geq2)
\end{array}\right\}\Rightarrow\nonumber\vspace{.2cm}\\
\beta_{0}=\frac{\alpha_{0}\left[\alpha_{-1}+\gamma_{1}
\right]}
{\beta_{1}-}
\frac{\alpha_{1}\gamma_{2}}{\beta_{2}-} \frac{\alpha_{2}\gamma_{3}}
{\beta_{3}-}\cdots,
\label{r2a}
\end{eqnarray}
\begin{eqnarray}
\left.
\begin{array}{l}
\alpha_{0}b_{1}+\left[\beta_{0}+\alpha_{-1}
\right]b_{0}=0,
\vspace{.2cm} \\
\alpha_{n}b_{n+1}+\beta_{n}b_{n}
+\gamma_{n}b_{n-1}=0\ (n\geq1)
\end{array}\right\}\Rightarrow\nonumber\vspace{.2cm}\\
\beta_{0}+\alpha_{-1}=\frac{\alpha_{0}\gamma_{1}}
{\beta_{1}-}\ \frac{\alpha_{1}\gamma_{2}}{\beta_{2}-}
 \frac{\alpha_{2}\gamma_{3}}
{\beta_{3}-}\cdots ,
\label{r3a}
\end{eqnarray}
where the coefficients $\alpha_n$, $\beta_n$ and $\gamma_n$
depend on the parameters of the equation and
on the summation index $n$ of the series.
The second and the third types of recurrence relations
occur only when the
parameters of the equation imply that $\alpha_{-1}\neq 0$,
as in the case of the Mathieu equation.

If $\gamma_{n}=0$ for some $n=N+1$,
where $N$ is a natural number or zero, the one-sided series
terminate at $n=N$ giving finite-series
solutions with $0\leq n\leq N$ \cite{arscott}.
In particular, for $N=0$ the series presents
just the first term except when the recurrence relations
are given by (\ref{r2a}): in this case the series
presents only the first term if $\alpha_{-1}+\gamma_{1}=0$.
Solutions given by finite series are also called quasi-polynomial
solutions or Heun polynomials.
On the other hand, the previous characteristic equations
were established by assuming that the summation
begins at $n=0$. However, the series begins at $n=N+1$
if $\alpha_{n}=0$ for some $n=N$. In this case we
must set $n= m+N+1$
and relabel the series coefficients in order
to get series beginning at $m=0$.

The fact that in each series
all the Bessel functions must be independent imposes
certain restrictions on the
parameters of the differential equation as in the
previous set of solutions where $B_{2}\neq 0,-1,-2,\cdots$
($Z_{m}^{(j)}$ and $Z_{-m}^{(j)}$ in general are
proportional to each other if $m$ is integer).
Such restrictions also assure that the
coefficients of the recurrence relations are well defined
in the sense that there are no vanishing denominators.
By transformations of variables which preserve
the form of equation (\ref{incegswe}) we
find new sets of solutions demanding different
restrictions on the parameters. In effect,
if $U(z)=U(B_{1},B_{2},B_{3};z_{0},q;z)$ denotes one
solution (or set of solutions) for equation
(\ref{incegswe}), then the transformations
$\mathscr{T}_1$, $\mathscr{T}_2$
and $\mathscr{T}_3$, given by
\begin{equation}
\mathscr{T}_{1}
U(z)=z^{1+({B_{1}}/{z_{0}})}\
U(C_{1},C_{2},C_{3};z_{0},q;z),\nonumber\hspace{1.7cm}
\end{equation}
\begin{equation}
 \mathscr{T}_{2}
U(z)=(z-z_{0})^{1-B_{2}-\frac{B_{1}}{z_{0}}} U(B_{1},D_{2},D_{3};
z_{0},q;z),\nonumber\hspace{.8cm}
\end{equation}
\begin{equation}\label{Transformacao3}
\mathscr{T}_{3}
U(z)=
U(-B_{1}-B_{2}z_{0},B_{2},
B_{3}-q z_{0};z_{0},-q;z_{0}-z),
\end{equation}
generate a group having eight (sets of) solutions. The constants
$C_{i}$ and $D_{i}$ are defined by
\begin{equation}
C_{1}=-B_{1}-2z_{0}, \quad
C_{2}=2+B_{2}+\frac{2B_{1}}{z_{0}},\hspace{2.1cm}
\nonumber
\end{equation}
\begin{equation}
C_{3}=B_{3}+
\left(1+\frac{B_{1}}{z_{0}}\right)
\left(B_{2}+\frac{B_{1}}{z_{0}}\right),\hspace{2.7cm}
\nonumber
\end{equation}
\begin{equation}\label{constantes-C-D}
D_{2}=2-B_{2}-\frac{2B_{1}}{z_{0}},\ 
D_{3}=B_{3}+
\frac{B_{1}}{z_{0}}\left[\frac{B_{1}}{z_{0}}
+B_{2}-1\right].
\end{equation}
The application of these rules is straightforward but
leads to solutions with different properties.

\subsection{Two subgroups of solutions in series of Bessel functions}

Given the initial set of solutions, $U_{1}^{(j)}$,
the rules $\mathscr{T}_1$
and $\mathscr{T}_2$ are used as
\begin{eqnarray}\label{sequence2}
U_{{1}}^{(j)}(z)
\stackrel {\mathscr{T}_1}{\longleftrightarrow}
U_{{2}}^{(j)}(z)
\stackrel {\mathscr{T}_2}{\longleftrightarrow}
U_{{3}}^{(j)}(z)
\stackrel {\mathscr{T}_1}{\longleftrightarrow}
U_{{4}}^{(j)}(z)
%
\end{eqnarray}
in order to generate a subgroup constituted by four
sets of solutions.
To get a second subgroup
of solutions, first we take
$U_{5}^{(j)}(z)=\mathscr{T}_{3}U_{1}^{(j)}(z) $
and, after this, we use the rules ${\mathscr{T}_1}$ and
${\mathscr{T}_2}$ in the same order as in
(\ref{sequence2}), that  is,
\begin{eqnarray}\label{sequence3}
U_{5}^{(j)}(z)
\stackrel {\mathscr{T}_1}{\longleftrightarrow}
U_{6}^{(j)}(z)
\stackrel {\mathscr{T}_2}{\longleftrightarrow}
U_{7}^{(j)}(z)
\stackrel {\mathscr{T}_1}{\longleftrightarrow}
U_{8}^{(j)}(z)
%
\end{eqnarray}
Then, we can check the following correspondence between
the solutions of the two subgroups
\letra
\begin{eqnarray}\label{t3-a}
U_{i}^{(j)}(z)\longleftrightarrow
U_{i+4}^{(j)}(z),\quad
(i=1,2,3,4)
\end{eqnarray}
in the sense that formally we have
\begin{eqnarray}\label{t3-b}
\beta_{n}^{(i)}=\beta_{n}^{(i+4)},\qquad
\alpha_{n}^{(i)}\ \gamma_{n+1}^{(i)}=
\alpha_{n}^{(i+4)}\ \gamma_{n+1}^{(i+4)}.
\end{eqnarray}
Besides this, if the summation
index $n$ takes the same values in both expansions,
the restrictions on the parameters,
the form of the recurrence relations and the order of the
Bessel functions are the same for both sets of solutions related
by  (\ref{t3-a}).
In this event both solutions have same characteristic
equation and their series coefficients are proportional to
each other.  However, in some cases one series breaks off on
the left while the other terminates on the right and so
these statements are not true, as in the example considered
in section III.

Next we write the first and the fifth set of solutions,
specify the restrictions on the parameters and the
conditions to use each of the preceding recurrence relations
(the other sets are in Appendix B). Thus,
%
%
\antiletra\letra
\begin{equation}
U_{1}^{(j)}(z) =z^{(1-B_{2})/2}
\displaystyle \sum_{n=0}^{\infty} (-1)^nb_{n}^{(1)}
Z_{2n+B_{2}-1}^{(j)}\left(2\sqrt{qz}\right),\nonumber\
\end{equation}
\begin{eqnarray}\label{bessel1b}
U_{5}^{(j)}(z)& =&(z-z_{0})^{(1-B_{2})/2}
\displaystyle \sum_{n=0}^{\infty} (-1)^nb_{n}^{(5)}\times\hspace{1.4cm}
\nonumber\\
& &\qquad Z_{2n+B_{2}-1}^{(j)}\left(2\sqrt{q(z-z_0)}\right),
\\
& &\hspace{3cm}[B_{2}\neq 0,-1,-2,\cdots].\nonumber
\end{eqnarray}
%
%
For $U_1^{(j)}$ we have \cite{eu}
\begin{equation}
\alpha_{n}^{(1)} = \frac{q z_{0}\left(n+1\right)
\left(n-\frac{B_{1}}{z_{0}}\right)}
{\left(n+\frac{B_{2}}{2}\right)\left(n+\frac{B_{2}}{2}+\frac{1}{2}
\right)},\hspace{2.7cm}
\nonumber
\end{equation}
\begin{eqnarray}
\beta_{n}^{(1)} &=&  4B_{3}-2q z_{0}
+4n\left(n+B_{2}-1\right)\nonumber\\
&-&\frac{2q z_{0}\left(\frac{B_{2}}{2}-1\right)
\left(\frac{B_{2}}{2}+\frac{B_{1}}{z_{0}}\right)}
{\left(n+\frac{B_{2}}{2}-1\right)
\left(n+\frac{B_{2}}{2}\right)},\hspace{1,8cm}\quad\nonumber
\end{eqnarray}
\begin{equation}\label{besse1c}
\gamma_{n}^{(1)}=
\frac{ q z_{0}\left(n+B_{2}-2\right)
\left(n+B_{2}+\frac{B_{1}}{z_{0}}-1\right)}
{\left(n+\frac{B_{2}}{2}-\frac{3}{2}\right)
\left(n+\frac{B_{2}}{2}-1\right)},
\end{equation}
in the recurrence relations for $b_{n}^{(1)}$ which are given by
\begin{eqnarray}\label{bessel1d}
 &&\text{ Eq. (\ref{r1a}) if }
B_{2}\neq 1,2; \quad \text{   Eq. (\ref{r2a}) if } B_{2}=1;\qquad
\nonumber\\
&&\text{   Eq. (\ref{r3a}) if }B_{2}=2.
\end{eqnarray}
For $U_5^{(j)}$ we get the following coefficients
($\beta_{n}^{(5)}  =\beta_{n}^{(1)}$)
\begin{equation*}
\alpha_{n}^{(5)}  =  -\frac{q z_{0}\left(n+1\right)
\left(n+B_2+\frac{B_{1}}{z_{0}}\right)}
{\left(n+\frac{B_{2}}{2}\right)\left(n+\frac{B_{2}}{2}+\frac{1}{2}
\right)},\nonumber\quad\
\end{equation*}
\begin{equation*}
\gamma_{n}^{(5)} =- \frac{ q z_{0}\left(n+B_{2}-2\right)
\left(n-1-\frac{B_{1}}{z_{0}}\right)}
{\left (n+\frac{B_{2}}{2}-\frac{3}{2}\right)
\left(n+\frac{B_{2}}{2}-1\right)},
\end{equation*}
in the recurrence relations (\ref{bessel1d}) for the
$b_{n}^{(5)}$. We find the formal relation
\antiletra
\begin{eqnarray}\label{pMathieu}
b_{n}^{(5)}=
\frac{(-1)^{n}\Gamma[n-(B_1/z_0)]}{\Gamma[n+B_2+(B_1/z_0)]}\
b_{n}^{(1)}.
\end{eqnarray}
Actually, this relation holds only if $(-B_{1}/z_0)$
and $(B_{2}+B_{1}/z_0)$  are not zero or negative integers;
on the contrary, one solution truncates on the left
($\alpha_n=0$ for some $n$) and the
other on the right ($\gamma_n=0$). There are similar relations
for the other solutions.

From relations (\ref{Bessel-assimptotico}), we get
\begin{eqnarray}
&&\lim_{z\rightarrow  \infty}U_{1}^{(3)}(z)\sim
e^{2i\sqrt{qz}}\ z^{(1/4)-(B_{2}/2)},\nonumber
\end{eqnarray}
\begin{eqnarray}\label{h1-h2}
&&\lim_{z\rightarrow  \infty}U_{1}^{(4)}(z)\sim
e^{- 2i\sqrt{qz}}\ z^{(1/4)-(B_{2}/2)},
\end{eqnarray}
that is, each solution presents one of the possible
behaviors given in equation (\ref{thome2}).
On the other side, equation (\ref{combinacao})
gives the solution $U_{1}^{(1)}$ as a linear combination
of $U_{1}^{(3)}$ and $U_{1}^{(4)}$
and, consequently, its behavior at
$z=\infty$ must be a linear combination of the
previous expressions. This implies that the solution
$U_{1}^{(1)}$ is bounded for $z=\infty$ only if both
$U_{1}^{(3)}$ and $U_{1}^{(4)}$ are bounded too.
The same holds for $U_{1}^{(2)}$.


The previous solutions lead to
the usual solutions in series of Bessel for Mathieu equation.
This equation has the form
\begin{eqnarray}\label{mathieu}
\frac{d^2w}{du^2}+\sigma^2\big[\mathrm{a}-2k^2\cos(2\sigma u)\big]w=0,
\quad \quad q=k^2,
\end{eqnarray}
where  $\sigma=1$ or $\sigma=i$ for the Mathieu or
modified Mathieu equation, respectively. Then,
by introducing
\begin{equation}
w(u)=U(z), \quad z=\cos^{2}(\sigma u),\nonumber
\end{equation}
\begin{equation}\label{Mathieu-equation}
z_{0}=1,\  B_{1}=-\frac{1}{2}, \ B_{2}=1, \
B_{3}=\frac{k^2}{2}-\frac{\mathrm{a}}{4}, \ q=k^2,\qquad
\end{equation}
into equation (\ref{incegswe}) we obtain the
above Mathieu equation. Using
these relations we can express the solutions of
the Mathieu equation in terms of trigonometric
(hyperbolic) functions. These solutions
are given by infinite series and the solutions
arising from the second subgroup have exactly
the same characteristic equations as the ones resulting
from the first subgroup. For example,
\letra
\begin{equation}
\displaystyle
w_{1}^{(j)}(u)=\sum_{n=0}^{\infty}(-1)^nb_{n}^{(1)}
Z_{2n}^{(j)}[2k\cos(\sigma u)],\nonumber
\end{equation}
\begin{equation}\label{mathieu-1}
\displaystyle
w_{5}^{(j)}(u)=\sum_{n=0}^{\infty}b_{n}^{(1)}
Z_{2n}^{(j)}[2ki\sin(\sigma u)],\ \
\end{equation}
with the recurrence relations
\begin{equation}
qb_{1}^{(1)}-\mathrm{a}b_{0}^{(1)}=0,\quad
qb_{2}^{(1)}+[4-\mathrm{a}]b_{1}^{(1)}+2qb_{0}^{(1)}=0,
\nonumber
\end{equation}
\begin{equation}\label{mathieu-2}
qb_{n+1}^{(1)}+\left[4n^2-\mathrm{a}\right]
b_{n}^{(1)}+qb_{n-1}^{(1)}=0, \ \
(n\geq 2),
\end{equation}
which follow from the second relation given in (\ref{bessel1d})
since $B_2=1$. Both sets have the same coefficients
because equation (\ref{pMathieu})
gives $b_{n}^{(5)}=(-1)^nb_{n}^{(1)}$.
According to the next section, the solutions $w_{1}^{(j)}$
converge for $|\cos(\sigma u)|\geq1$ whereas $w_{5}^{(j)}$
converge only for $|\sin(\sigma u)|\geq1$ if $j=2,3,4$.
{Despite this, in a common domain of convergence
$w_{1}^{(j)}$ is a constant multiple of $w_{5}^{(j)}$,
except if $k=0$  \cite{arscott,McLachlan}.} 
%

For the general case, we have found no proof for the equivalence
between the sets of solutions of the first and second subgroups.
In fact, equivalence
could take place only  between two sets of infinite-series solutions where
the summations begin at $n=0$. For instance,
for the first and fifth sets, this requires that neither $(-B_1/z_0)$
nor $B_2+(B_1/z_0)$ are zero or negative integer since
this implies infinite series in both sets.

On the other side, some sets
of solutions admit the limit $z_0\rightarrow0$,
in which case equation (\ref{incegswe})
becomes a double-confluent Heun equation with
two (irregular) singularities located at $z=0$ and
$z=\infty$. In this case the second subgroup
is irrelevant and only
the sets $U_{1}^{(j)}(z)$ and $U_{3}^{(j)}(z)$
admit such limit, as we can see by examining the
recurrence relations for the series coefficients.
\subsection{Convergence of the solutions}
Now the ratio test is used to get the convergence of the
first set of solutions. For the other sets the convergence follows
from the transformation rules. The following convergence
regions refer to one-sided
series; these series hold only if the differential equation
has an arbitrary parameter that allows to satisfy the
characteristic equations. If there is no free
parameter, it is possible to obtain convergent two-sided
series solutions (summation extending from negative to
positive infinity) by introducing a characteristic
parameter $\nu$ into the solutions but,
then, the domains of convergence are changed.

For $n\rightarrow\infty$ the  recurrence relations
for $b_{n}^{(1)}$ become (we write $b_{n}$
instead of $b_{n}^{(1)}$)
\begin{eqnarray*}
\displaystyle &&qz_0\left[1-\frac{1}{n}\left(B_2+\frac{B_1}{z_0}-\frac{1}{2}\right)+
O\left(\frac{1}{n^2}\right)\right]\frac{{b}_{n+1}}{{b}_{n}}+
\nonumber\\
&&\big[ 4n(n+B_2-1)+O\left(1\right)\big]+ \vspace{1.5mm}\\
&&
qz_0\left[1+\frac{1}{n}\left(B_2+\frac{B_1}{z_0}-
\frac{1}{2}\right)+O\left(\frac{1}{n^2}\right)\right]
\frac{{b}_{n-1}}{{b}_{n}}=0,
\end{eqnarray*}
whose minimal solution for $z_0\neq 0$ is
\antiletra\letra
\begin{eqnarray}\label{minimal}
\displaystyle
\frac{b_{n+1}}{b_{n}}\sim
-\frac{qz_0}{4n^2}\left[1+\frac{1}{n}\left(\frac{B_1}{z_0}-\frac{3}{2}\right)
+O\left( \frac{1}{n^2}\right)  \right]\Rightarrow\nonumber\vspace{1.5mm}\\
\displaystyle
\frac{b_{n-1}}{b_{n}}\sim
-\frac{4n^2}{qz_0}\left[ 1-\frac{1}{n}\left(\frac{B_1}{z_0}+\frac{1}{2}\right)+
O\left(\frac{1}{n^2} \right) \right],\quad
\end{eqnarray}
where the notation having the form $f_n\sim g_n$ means
that $g_n$ is an asymptotic approximation to $f_n$
as $n\rightarrow \infty$, that is, $f_n/g_n=1$  when
$n\rightarrow \infty$ \cite{olver}. Similarly, for $z_0=0$
\begin{eqnarray}\label{minimal2}
\displaystyle
\frac{b_{n+1}}{b_{n}}\sim
-\frac{qB_1}{4n^3}\left[1
+O\left( \frac{1}{n}\right)  \right]\Rightarrow\nonumber\\
\frac{b_{n-1}}{b_{n}}\sim
-\frac{4n^3}{qB_1}\left[1+ O\left( \frac{1}{n}\right) \right],\quad
\end{eqnarray}

Firstly we regard the expansions in series of Bessel functions of the
first kind. Using the first relation given in equation (\ref{JY}), we find
\[\frac{
J_{2n+B_2+1}\left(2\sqrt{qz}\right)}
{J_{2n+B_2-1}\left(2\sqrt{qz}\right)}
\sim\frac{qz}{4n^2}\left[1+O\left(\frac{1}{n} \right) \right].\]
Then, keeping only the leading terms, we obtain
\begin{equation*}
\frac{b_{n+1}\
J_{2n+B_2+1}\left(2\sqrt{qz}\right)}
{b_{n}\ J_{2n+B_2-1}\left(2\sqrt{qz}\right)} \sim\
\frac{z_0q^2  z}{16n^4},\ (z_0\neq 0)
\end{equation*}
\begin{equation*}
\frac{b_{n+1}\ J_{2n+B_2+1}\left(2\sqrt{qz}\right)}
{b_{n}\ J_{2n+B_2-1}\left(2\sqrt{qz}\right)} \sim\
\frac{B_1q^2  z}{16n^5},\ (z_0=0).
\end{equation*}
Therefore, the series converges for
any finite $z$, but  the ratio test is
inconclusive when $z\rightarrow\infty$. For
this case we proceed as in the case of the
Mathieu equation \cite{arscott}. Thus, by using
equation (\ref{Bessel-assimptotico}) we write
($z\rightarrow\infty$)
\begin{eqnarray*}
 &&J_{2n+B_1-1}(2\sqrt{qz})\sim
(-1)^n\left(\frac{1}{\sqrt{qz}}\right)^{\frac{1}{2}}\times\nonumber\\
&&\cos\left[2\sqrt{qz}-\frac{1}{2}(B_2-1)\pi-\frac{1}{4}\pi\right],
 \ |\arg \sqrt{qz}|<\pi.
\end{eqnarray*}
By inserting this into $U_1^{(1)}$, we find
\antiletra
\begin{eqnarray}
&&\lim_{z\rightarrow\infty}U_{1}^{(1)}(z)
=
z^{(1/4)-(B_2/2)}\times\nonumber\\
&&\cos\left[2\sqrt{qz}-\frac{1}{2}(B_2-1)\pi-\frac{1}{4}\pi\right]
\displaystyle \sum_{n=0}^{\infty} b_{n}^{(1)}.
\end{eqnarray}
Since the series
$\sum_{n=0}^{\infty} b_{n}^{(1)}$ converges,
the solution $U_{1}^{(1)}(z)$ also converges at $z=\infty$.
However, if the cosine is expressed in terms of exponential functions,
we obtain a linear combination of asymptotic behaviors
of the expansion in Hankel functions, as observed after equation
(\ref{h1-h2}). Thus, depending on the values
of $B_2$ and $|\arg (\sqrt{qz)}|$, the solution
$U_{1}^{(1)}$ may be bounded or unbounded at $z=\infty$.

Now we consider the expansions in series
of Hankel functions and Bessel functions of
the second kind. For $\alpha=2n+B_2-1$
and $x=2\sqrt{qz}$, relations (\ref{JY}) give
\begin{eqnarray*}
&&\frac{
Z_{2n+B_2+1}^{(j)}\left[2\sqrt{qz}\right]}
{Z_{2n+B_2-1}^{(j)}\left[2\sqrt{qz}\right]}
\sim\\
&& \frac{4n^2}{qz}\left[1+\frac{2B_2-1}{2n}+
\frac{B_2[B_2-1]}{n^2} \right],\quad
\left[  j=2,3,4\right] .
\end{eqnarray*}
Combining this with the results given in equations (\ref{minimal})
and  (\ref{minimal2}), we find
\begin{eqnarray*}
&&\displaystyle
\frac{b_{n+1}
Z_{2n+B_2+1}^{(j)}\left(2\sqrt{qz}\right)}
{b_{n}Z_{2n+B_2-1}^{(j)}\left(2\sqrt{qz}\right)}
\sim\\
&&\quad-\frac{z_0}{z}\left[1+\frac{1}{n}\left(B_2-2+\frac{B_1}{z_0}\right)+
O\left(\frac{1}{n^2}\right)  \right],
\ z_0\neq 0,
\end{eqnarray*}
\begin{eqnarray*}
&&\displaystyle
\frac{b_{n+1} Z_{2n+B_2+1}^{(j)}\left(2\sqrt{qz}\right)}
{b_{n} Z_{2n+B_2-1}^{(j)}\left(2\sqrt{qz}\right)}\sim -
\frac{B_1}{nz}\left[ 1+O\left( \frac{1}{n}\right) \right] ,
z_0=0.
\end{eqnarray*}
%
Then, the ratio test implies that the series in the
solutions $U_{1}^{(j)}$ ($j=2,3,4$) converge for
$|z|>0$ if $z_0=0$. On the other hand, for $z_0\neq 0$
we have ($n\rightarrow\infty$)
\begin{eqnarray}
&\displaystyle\vline\frac{b_{n+1}
Z_{2n+B_2+1}^{(j)}\left(2\sqrt{qz}\right)}
{b_{n}Z_{2n+B_2-1}^{(j)}\left(2\sqrt{qz}\right)}\vline
\displaystyle=\frac{|z_0|}{|z|}\times\nonumber\\
&\displaystyle\left[1+\frac{1}{n}\text{Re}
\left(B_2-2+\frac{B_1}{z_0}\right)+O\left( \frac{1}{n^2}
\right)\right], (z_0\neq 0).\quad
\end{eqnarray}
Thus, if  $z_0\neq 0$ in general the series converge for
$|z|>|z_0|$ since the right side of this equation is
$<1$ but, by the Raabe test \cite{watson,knopp}, they converge
absolutely also for $|z|=|z_0|$ if the
numerator of $n$ is $< -1$, that is, if $\text{Re}(B_2+B_1/z_0)<1$.
This possibility was not noticed before  \cite{eu}
because the term of order $1/n$ was not considered.

The convergence for the other sets of solutions
are obtained by applying the transformations
rules to $U_{1}^{(j)}$.  For $j=2,3,4$, we find that
these solutions converge for $|z|>|z_0|$ in
the first subgroup, and  for $|z-z_0|>|z_0|$ in the
second subgroup. The special cases are given by:
{
\letra
\begin{equation}\label{convergencia1}
|z|\geq|z_0| \text{ if }
\begin{cases}\text{Re}\left(  B_2+\frac{B_1}{z_0}\right) <1:
U_{1}^{(j)} \text{ and }
U_{2}^{(j)},\vspace{2mm} \\
\text{Re}\left(  B_2+\frac{B_1}{z_0}\right) >1:
U_{3}^{(j)} \text{ and }
U_{4}^{(j)};
\end{cases}
\end{equation}
\begin{equation}\label{convergencia2}
|z-z_0|\geq|z_0| \text{ if }
\begin{cases}\text{Re}\left( \frac{B_1}{z_0}\right) >1:
U_{5}^{(j)} \text{ and }
U_{6}^{(j)},\vspace{2mm} \\
\text{Re}\left( \frac{B_1}{z_0}\right) <1:
U_{7}^{(j)} \text{ and }
U_{8}^{(j)}.
\end{cases}
\end{equation}
}
The above conditions have been derived for infinite
series where the summation begins at $n=0$,
as in the case of the Mathieu equation. A counterexample
occurs in the analysis of the inverted potential, as
explained after  the paragraph containing the
solutions (\ref{infintas-3}).
%
%
\section{The inverted potential}
In this section we get solutions for
equation (\ref{cho-ho}) of the inverted potential by
using the preceding solutions in series of Bessel
functions of the first kind for the Whittaker-Ince limit of the CHE.
We will find one pair of finite-series solutions and two pairs of
infinite-series solutions, all of them convergent
and bounded for any value of the variable $u$. The
two pairs of infinite series have the same characteristic
equation, but only the solutions of one pair can be
expressed as a linear combination of two series of
Hankel functions converging for all values of the
independent variable.

The substitutions
\antiletra
\begin{eqnarray}\label{vetor1}
\psi(u)=\big[\cosh{u}\big]^{-l+\frac{1}{2}}U(z),\ z=-\sinh^2u,
\end{eqnarray}
bring equation (\ref{cho-ho}) to the form
\begin{eqnarray}\label{teller}
&&z(z-1)\frac{d^{2}U}{dz^{2}}+\left[ -\frac{1}{2}+\left( \frac{3}{2}-l\right) z\right]
\frac{dU}{dz}+\nonumber\\
&&\left[\frac{{\cal E}}{4}-\frac{\text{b}^2}{16}+
\left(\frac{l}{2}-\frac{1}{4} \right)^2 -\frac{\text{b}^2}{16}(z-1)\right]U=0,
\end{eqnarray}
which is Whittaker-Ince limit (\ref{incegswe}) of the CHE
with
\begin{eqnarray}
&&z_0=1,\qquad B_{1}=-\frac{1}{2}, \qquad B_{2}=\frac{3}{2}-l,
\nonumber\
\end{eqnarray}
\begin{eqnarray}\label{parameters-ho}
 &&B_{3}=\frac{{\cal E}}{4}-\frac{\text{b}^2}{16} +
\left(\frac{l}{2}-\frac{1}{4}\right)^{2}, \quad q=-\frac{\text{b}^{2}}{16}.
\end{eqnarray}
Thus, the Schr\"odinger equation (\ref{cho-ho}) will be
solved by replacing $U(z)$ in equation (\ref{vetor1})
by the the expansions in series of Bessel functions
of the first kind with the parameters (\ref{parameters-ho}).
Notice that for $\text{b}=0$, the potential $V(u)$
reduces to a hyperbolic P\"{o}schl-Teller potential
whose solutions are given by hypergeometric functions \cite{flugge}.

In section III.A we write down the pair of finite-series solutions
which allow to get the `quasi-solvable' part of the
energy spectrum. We prove that these are
degenerate for any finite value of the parameter $l$ and
adapt a method devised by Bender and Dunne \cite{bender}
(for a potential which leads to a biconfluent Heun equation)
to find the energy levels
and the series coefficients. In section III.B we discuss
the infinite-series solutions. In all cases, using
the  parameters (\ref{parameters-ho}) we
find that the order of
the Bessel functions is half-integer
and so these functions are
represented by finite series rather than by infinite
ones \cite{abramowitz}. Besides this, the recurrence
relations are always given by equation (\ref{r1a})
since $\alpha_{-1}^{(i)}=0$.

\subsection{Finite-series solutions}
According to section II.A, if $\gamma_{n=N+1}=0$
in the recurrence relations (\ref{r1a}), the series
terminates at $n=N$ and we obtain a finite-series solution.
In this case the recurrence relations can be written in the form
\begin{eqnarray}
\label{matriz}
\left(
\begin{array}{cccccccc}
\beta_{0} & \alpha_{0} &0   &\cdots  &       &  &   &   0    \\
\gamma_{1}&\beta_{1}   & \alpha_{1} &         &
&  &  \\
   0    &\gamma_{2} & \beta_{2}    &\alpha_{2}&   &   &     &    \\
 \vdots  &         &            &           &   &  &          \\
        &   &   &   &    &  \gamma_{N-1}& \beta_{N-1}&\alpha_{N-1}\\
  0  &\cdots    &   &   &       &       0    &\gamma_{N} & \beta_{N}
\end{array}
\right) \left(\begin{array}{l}
b_{0}  \\
b_{1} \\
b_{2} \\
 \vdots   \\
b_{N-1}\\
b_{N}
\end{array}
\right)\nonumber\\
=0.\quad
\end{eqnarray}
This system has nontrivial solutions only if the determinant of the
above tridiagonal matrix vanishes. Besides this,
if (as in the present problem) the elements of this matrix
are real and
\begin{eqnarray}\label{lemma}
\alpha_{i}\gamma_{i+1}>0, \ \  \ \ 0\leq i\leq N-1,
\end{eqnarray}
then the $N+1$ roots of the determinant are real and distinct
\cite{arscott}.

Only $U_1^{(j)}$ and $U_2^{(j)}$ give finite-series wavefunctions.
In order to prove the degeneracy and apply the procedure of
Bender and Dunne, we redefine
the series coefficients as
\begin{eqnarray*}
&&b_{n}^{(1)}=\frac{\Gamma \left[ 2n+B_2\right]\
{P}_{n} }{(-qz_0)^{n}\ n!\
\Gamma\left[ n-({B_{1}}/{z_{0}})\right] }
,\qquad
\end{eqnarray*}
\begin{eqnarray*}
&&b_{n}^{(2)}=\frac{\Gamma \left[ 2n+2+B_2
+({2B_{1}}/{z_{0}})\right]\ Q_{n} }{(-qz_0)^{n}\ n!\
\Gamma\left[ n+2+({B_{1}}/{z_{0}})\right] }.
\end{eqnarray*}
Then, inserting the solution $U_1^{(1)}$
and  $U_2^{(1)}$ into equation (\ref{vetor1}) and
using the parameters given in (\ref{parameters-ho}),
we get the following pair
($\psi_{1}^{\text{e}},\psi_{1}^{\text{o}}$) of even and odd
solutions
\begin{eqnarray}
&&\psi_{1}^{\text{e}}(u)=\big[\tanh{u}\big]^{l-\frac{1}{2}}
\displaystyle \sum_{n=0}^{l-1}(-1)^n
\left(\frac{4}{\text{b}} \right)^{2n}\times\nonumber\\
&&\frac{\Gamma \left[2 n-l+({3}/{2})\right]{P}_{n} }
{n!\
\Gamma\left[ n+({1}/{2})\right] }
  \displaystyle
  J_{2n-l+({1}/{2})}\left(\frac{\text{b}}{2}\sinh{u}\right),
\nonumber
\end{eqnarray}
\begin{eqnarray}\label{primeiro-psiA}
&&\psi_{1}^{\text{o}}(u) =\big[\tanh{u}\big]^{l-\frac{1}{2}}
\displaystyle \sum_{n=0}^{l-1}(-1)^n
\left(\frac{4}{\text{b}} \right)^{2n} \times\nonumber\\
&&\frac{\Gamma \left[2 n-l+({5}/{2})\right]{Q}_{n} }
{n!\
\Gamma\left[ n+({3}/{2})\right] }
  \displaystyle
J_{2n-l+({3}/{2})}\left(\frac{\text{b}}{2}\sinh{u}\right),
\end{eqnarray}
where the recurrence relations for the coefficients $P_n$ are
\letra
\begin{eqnarray}\label{recP}
{P}_{n+1}+{\beta}_{n}{P}_{n}+
{\gamma}_{n}{P}_{n-1}=0,\qquad (P_{-1}=0)
\end{eqnarray}
with
\begin{eqnarray}
{\beta}_{n}&=&-E_{l}-n\left(n-l+\frac{1}{2}\right)
\nonumber\\
&-&
\frac{\text{b}^{2}(2l+1)(2l-1)}{32(4n-1-2l)(4n+3-2l)},
\hspace{2.5cm}
\nonumber
\end{eqnarray}
\begin{eqnarray}\label{El}
{\gamma}_{n}=\frac{\text{b}^4}{64}
\frac{\ n(2n-1)(2n-2l-1)(n-l)}
{(4n+1-2l)(4n-3-2l)
(4n-1-2l)^{2}},\ \ \\ 
\left[ E_{l}:=\frac{{\cal E}}{4}-\frac{\text{b}^2}{32} +\left(
\frac{l}{2}-\frac{1}{4}\right)^{2}\right]  .\nonumber
\end{eqnarray}
The recurrence relations for $Q_{n}$ are
\antiletra\letra
\begin{eqnarray}\label{recQ}
{Q}_{n+1}+\tilde{\beta}_{n}{Q}_{n}+
\tilde{\gamma}_{n}{Q}_{n-1}=0,\quad (Q_{-1}=0)
\end{eqnarray}
with
\begin{eqnarray}
\tilde{\beta}_{n}&=&-E_{l}-\left(n-l+1\right)\left(n+\frac{1}{2}\right)
\nonumber\\
&-&
\frac{\text{b}^{2}(2l+1)(2l-1)}{32(4n+1-2l)(4n+5-2l)},
\nonumber\hspace{2cm}
\end{eqnarray}
\begin{equation}\label{recQ-b}
\tilde{\gamma}_{n}=\frac{\text{b}^4}{64}
\frac{ n(2n+1)(2n-2l+1)(n-l)}
{(4n-1-2l)(4n+3-2l)
(4n+1-2l)^{2}}.
\end{equation}
   From the relation $J_{\lambda}\left(-x\right)$ $=$
$(-1)^{\lambda} J_{\lambda}\left( x\right)$,
it turns out that these solutions are even and odd,
that is, $\psi_{1}^{\text{e}}(-u)$ $=$
$\psi_{1}^{\text{e}}(u)$ and $\psi_{1}^{\text{o}}(-u)$ $=$
$-\psi_{1}^{\text{o}}(u)$.

The series are finite
because the coefficients $\gamma_n$ and $\tilde{\gamma}_n$
of $P_{n-1}$ and $Q_{n-1}$ vanish
for $n=l$ and, consequently, the series terminate at
$n=l-1$ as stated above. In virtue of equation (\ref{lemma}),
each eigenfunction corresponds to $l$ distinct and
real eigenvalues. In addition, from $J_{\lambda}(x)= (x/2)^{\lambda}/\Gamma(\lambda+1)$
when $x\rightarrow 0$, we find
\begin{eqnarray*}
\lim_{u\rightarrow{0}}\psi_{1}^{\text{e}}(u)
\sim\big[\cosh{u}\big]^{-l+\frac{1}{2}}
\rightarrow\text{finite},
\end{eqnarray*}
and from the first  of equations (\ref{Bessel-assimptotico})  we get
\begin{eqnarray*}
\displaystyle\lim_{u\rightarrow{\pm \infty}}
\psi_{1}^{\text{e}}(u)&\sim&
\frac{[\tanh{u}]^{l-\frac{1}{2}}}{\sqrt{\sinh{u}}}
\cos\left[
\frac{\text{b}}{2}\sinh{u}+\frac{1}{2}(l-1)\pi\right]
\vspace{2mm}\\
& \times&\displaystyle\sum_{n=0}^{l-1}
\left(\frac{4}{\text{b}} \right)^{2n}
\frac{\Gamma \left[2 n-l+{3}/{2}\right]{P}_{n} }
{n!\
\Gamma\left[ n+{1}/{2}\right] }
\rightarrow 0.
\end{eqnarray*}
Thence,  $\psi_{1}^{\text{e}}(u)$
is bounded also at the singular points of the
equation ($z=0$ and $z=-\infty$). The same
holds for the solutions $\psi_{1}^{\text{o}}(u)$.
In fact, expansions in series of the other Bessel
functions are also bounded for all values of $u$
but do not present definite parity.

The degeneracy of the previous solutions is established by arranging
the recurrence relations in the matrix form (\ref{matriz}).
For $P_{n}$ we write $\mathbb{A}\vec{P}=0$, where
\antiletra
\[\vec{P}= (P_0, P_1,\cdots, P_{l-1})^t\]
($t$ means `transpose') and
\begin{eqnarray}\label{matrizP}
\mathbb{A}=\left(
\begin{array}{ccccccc}
\beta_{0} & 1 &   &     &      &   &   0    \\
\gamma_{1}&\beta_{1}   & 1 &            &
 &  \\
    &\gamma_{2} & \beta_{2}    &. &  &  &   \\
&   &  . &   &   &   &     \\
        &   &   &      & \gamma_{l-2}& \beta_{l-2}&1\\
  0  &    &   &   &     0     &\gamma_{l-1} & \beta_{l-1}
\end{array}
\right).
\end{eqnarray}
For $Q_n $ we write $\mathbb{B}\vec{Q}=0$, where
\[\vec{Q}=
 (Q_0, Q_1,\cdots, Q_{l-1})^t\]
and
\begin{eqnarray*}
\mathbb{B}=
\left(
\begin{array}{ccccccc}
\tilde{\beta}_{0} & 1 &   &     &      &   &   0    \\
\tilde{\gamma}_{1}&\tilde{\beta}_{1}   & 1 &            &
 &  \\
    &\tilde{\gamma}_{2} & \tilde{\beta}_{2}    &. &  &  &   \\
&   &  . &   &   &   &     \\
        &   &   &      & \tilde{\gamma}_{l-2}& \tilde{\beta}_{l-2}&1\\
  0  &    &   &   &          &\tilde{\gamma}_{l-1} & \tilde{\beta}_{l-1}
\end{array}
\right).
\end{eqnarray*}
This matrix can be rewritten as
\begin{eqnarray}\label{matrizQ}
\mathbb{B}=\left(
\begin{array}{ccccccc}
\beta_{l-1} & 1 &   &     &      &   &   0    \\
\gamma_{l-1}&\beta_{l-2}   & 1 &    & &  &  \\
    &\gamma_{l-2} & \beta_{l-3}    &.&  &  &   \\
&   &  . &   &   &     & \\
        &   &   &      & \gamma_{2}& \beta_{1}&1\\
  0  &    &   &   &    0    &\gamma_{1} & \beta_{0}
\end{array}
\right),
\end{eqnarray}
due to the identities
\begin{eqnarray*}
\begin{array}{llll}
\tilde\beta_{0}=\beta_{l-1},&
\tilde\beta_{1}=\beta_{l-2},& \tilde\beta_{2}=\beta_{l-3},&
\cdots, \quad\tilde\beta_{l-1}=\beta_{0},
\end{array}\end{eqnarray*}
\begin{eqnarray*}
\begin{array}{llll}
\tilde{\gamma}_{1}=\gamma_{l-1},&
\tilde{\gamma}_{2}=\gamma_{l-2},&
\tilde\gamma_{3}=\gamma_{l-3},&
\cdots, \quad\tilde\gamma_{l-1}=\gamma_{1}.
\end{array}
\end{eqnarray*}
Therefore, the two matrices are constituted by the same elements.
To prove the degeneracy, it is sufficient
to show that these matrices possess
the same roots, that is, $\det{\mathbb{A}}=$
$\det{\mathbb{B}}$. This is obvious for $l=1$ and $l=2$.
For $l\geq 3$, we use the
$l\text{-by-}l$ antidiagonal matrix
$\mathbb{S}$ having $1$'s on
the antidiagonal as the only nonzero elements, that is,
\begin{eqnarray}
\mathbb{S}=\mathbb{S}^{-1}=\left(
\begin{array}{ccc}
  &   &   1    \\
            & . &  \\
1 & &
\end{array}
\right),\qquad \det{\mathbb{S}}=-1.
\end{eqnarray}
Then we find the similarity relation
\begin{eqnarray}
\mathbb{A}=\mathbb{S}^{-1}\mathbb{B}^{t}\mathbb{S}=
\mathbb{S}\mathbb{B}^{t}\mathbb{S},
\end{eqnarray}
where $\mathbb{B}^{t}$ is the transpose
of $\mathbb{B}$.
Thus, from the properties of the
determinants it follows that
$\det{\mathbb{A}}=\det{\mathbb{B}}$ and, therefore,
the finite-series solutions are degenerate for any finite $l$.
On the other hand, the eigenvalues may be computed
by equating to zero the
determinants of the preceding matrices.  However, the
procedure of Bender and Dunne gives as well the
coefficients $P_{n}$ and $Q_n$ as polynomials
of degree $n$ in the the parameter $E_{l}$.
The procedure is implemented by taking $P_{0}=Q_{0}=1$
as initial conditions and by using the recurrence relations
to generate the other coefficients.
For a fixed $l$, the eigenvalues
are obtained by requiring
that $P_{l}=0$ or $Q_{l}=0$, since the
series terminate at $n=l-1$.
Thus, equations (\ref{recP}) and (\ref{El}) yield
\letra
\begin{eqnarray}\label{factorisation}
P_{n+1}=[E_{l}+k_{nl}]P_{n}-\gamma_{n}P_{n-1},
\  P_{-1}=0, \ P_{0}=1\ \
\end{eqnarray}
with
\begin{equation}
k_{nl}=n\left(n-l+\frac{1}{2}\right)+
\frac{\text{b}^{2}(2l+1)(2l-1)}{32(4n-1-2l)(4n+3-2l)},
\end{equation}
wherefrom we find
\antiletra
\begin{eqnarray}
P_{0}=1,\qquad
P_{1}= {E}_{l}+k_{0l},\nonumber
\end{eqnarray}
\begin{eqnarray*}
&&P_{2}={E}_{l}^2+[k_{0l}+k_{1l}] {E}_{l}+
k_{0l}k_{1l}-\gamma_{1},
\end{eqnarray*}
\begin{eqnarray}
&&P_{3}={E}_{l}^3+
\big[k_{0l}+k_{1l}+ k_{2l}\big]
{E}_{l}^2+
\big[k_{0l}k_{1l}+k_{0l}k_{2l}\nonumber\\
&&
+k_{1l}k_{2l}-\gamma_{1}-\gamma_2\big]{E}_{l}+
k_{2l}\big[k_{0l}k_{1l}-\gamma_{1}\big]-
k_{0l}\gamma_{2},\qquad
\end{eqnarray}
and so on.  Similarly, we write the relations (\ref{recQ})
and (\ref{recQ-b}) for $Q_{n}$ as
\letra
\begin{equation}
Q_{n+1}=[E_{l}+\tilde{k}_{nl}]Q_{n}-
\tilde{\gamma}_{n}Q_{n-1},
\ Q_{-1}=0, \ Q_{0}=1,
\end{equation}
where
\begin{equation}
\tilde{k}_{nl}=\left[n-l+1\right]\left[n+\frac{1}{2}\right]
+
\frac{\text{b}^{2}(2l+1)(2l-1)}{32(4n+1-2l)(4n+5-2l)}.
\end{equation}
Then, the expressions for $Q_{n}$ are obtained by replacing
$k_{nl}$ and $\gamma_{n}$ by
$\tilde{k}_{nl}$ and $\tilde{\gamma}_{n}$
in the expressions for $P_{n}$.

As an example we find the energies and the respective
eigenfunctions for $l=1$ and $l=2$. For $l=1$ the
energy that follows from the
condition $P_1=Q_1=0$ is
\antiletra\letra
\begin{eqnarray}
E_{l=1}-\frac{\text{b}^2}{32}=0\quad \Rightarrow\quad {\cal E}=
\frac{1}{4}\left( \text{b}^2-1\right),
\end{eqnarray}
corresponding to the degenerate pair of eigenfunctions
\begin{eqnarray}
\ \psi_{1}^{\text{e}}(u) =
\sqrt{\tanh{u}}\ J_{-\frac{1}{2}}\left( \frac{\text{b}}{2}\sinh{u}\right),
\nonumber
\end{eqnarray}
\begin{eqnarray}
\psi_{1}^{\text{o}}(u)=
\sqrt{\tanh{u}}\ J_{\frac{1}{2}}\left( \frac{\text{b}}{2}\sinh{u}\right).
\end{eqnarray}
For $l=2$ the condition
$P_2=Q_2=0$ leads to
\antiletra\letra
\begin{eqnarray}
&&E_{l=2}^{\pm}=\frac{1}{4}
\left[ \frac{\text{b}^2}{8}+1\pm\sqrt{1+\text{b}^2}\right]
\quad \Rightarrow\nonumber\\
 &&{\cal E^{\pm}}=\frac{1}{4}
\left[ \text{b}^2-5\right] \pm\sqrt{1+\text{b}^2}.
\end{eqnarray}
Since, $P_1=$ $E_2+(3\text{b}^2/32)$ and
$Q_1=$ $E_2-(1/2)-(5\text{b}^2/32)$ these
energies are associated with the eigenstates
\begin{eqnarray}
\ \ \psi_{1}^{\text{e}\pm}(u)&=&\left( \tanh{u}\right) ^{\frac{3}{2}}
\bigg[J_{-\frac{3}{2}}\left( \frac{\text{b}}{2}\sinh{u}\right)
\nonumber\\
& +&\frac{2}{\text{b}^2}
\left(\frac{\text{b}^2}{2}+
1\pm\sqrt{\text{b}^2+1} \right) J_{\frac{1}{2}}\left( \frac{\text{b}}{2}\sinh{u}\right) \bigg], \nonumber
\end{eqnarray}
\begin{eqnarray}
\psi_{1}^{\text{o}\pm}(u) =(\tanh{u})^{\frac{3}{2}}
\bigg[ J_{-\frac{1}{2}}\left( \frac{\text{b}}{2}\sinh{u}\right)+\hspace{1.4cm}
\nonumber\\
\frac{2}{\text{b}^2}
\left(\frac{\text{b}^2}{2}+1\mp\sqrt{\text{b}^2+1} \right)
J_{\frac{3}{2}}\left( \frac{\text{b}}{2}
\sinh{u}\right) \bigg],\
\end{eqnarray}
up to normalization factors.

These energies for $l=1,2$ are the same
found by Cho and Ho  \cite{cho-2}.
However, the eigenfunctions differ from theirs,
as we can see by writing the Bessel functions
in terms of elementary functions
via the formulas written in Appendix A.
Indeed, their solutions may be obtained from
another group of expansions in series of Bessel functions
given in \cite{lea}.

Notice that the coefficients of the preceding finite series
factorize in the same manner as the coefficients of
the problem considered by Bender and Dunne
\cite{bender}, that is, for a fixed $l$ one has
\antiletra
\begin{eqnarray}
P_{l+i}(E_l)=p_{i}(E_l) P_{l}(E_l), \nonumber\hspace{1.35cm}
\end{eqnarray}
\begin{eqnarray}\label{alhaidari}
Q_{l+i}(E_l)=
q_{i}(E_l)Q_{l}(E_l),
\ (i\geq 0)\end{eqnarray}
where $p_i$ and $q_i$ are polynomials of degree $i$ in $E_l$.
For example, by taking $n=l$ ($\gamma_l=0$) and $n=l+1$
in equation (\ref{factorisation}), we find
\begin{eqnarray*}
P_{l+1}&=&(E_l+k_{l,l})P_l=p_1P_l ,
\hspace{3.2cm}
\end{eqnarray*}
\begin{eqnarray*}
P_{l+2}&=&[(E_l+k_{l+1,l})(E_l+k_{l,l})-\gamma_{l+1}]P_l=p_2P_l ,
\end{eqnarray*}
and, by induction, we obtain the previous expression for $P_{l+i}$.

\subsection{Infinite-series solutions}
The solutions $U_{3}^{(1)}$ and $U_{4}^{(1)}$ lead
respectively to odd and even infinite-series
wavefunctions which are bounded for any value of $u$,
namely,
\letra
\begin{eqnarray}
\psi_{2}^{o}(u) =\big[\tanh{u}\big]^{-l-\frac{1}{2}}
\displaystyle \sum_{n=0}^{\infty}(-1)^n
b_{n}^{(3)}\times\nonumber\\
 J_{2n+l+({3}/{2})}\left(\frac{\text{b}}{2}\sinh{u}\right),
\nonumber
\end{eqnarray}
\begin{eqnarray}\label{primeiro-psiB}
\psi_{2}^{\text{e}}(u) =\big[\tanh{u}\big]^{-l-\frac{1}{2}}
\displaystyle \sum_{n=0}^{\infty}(-1)^n b_{n}^{(4)}\times\nonumber\\
 J_{2n+l+({1}/{2})}\left(\frac{\text{b}}{2}\sinh{u}\right) .
\end{eqnarray}
In the recurrence relations (\ref{r1a}) for $b_{n}^{(3)}$
the coefficients are
\begin{equation}
\alpha_{n}^{ (3)} = -
 \frac{\text{b}^2(n+1)(2n+3)}
{2\left(4n+2l+5\right)\left(4n+2l+7\right)},\nonumber
\hspace{2.2cm}
\end{equation}
\begin{eqnarray}
\beta_{n}^{ (3)}  &= &{\cal E}-
\frac{\text{b}^2}{8}+\left( l-\frac{1}{2}\right)^2 +
4(n+1)\left(n+l+\frac{1}{2}\right)\nonumber\\
&+&\frac{\text{b}^2\left(2l+1\right)
\left(2l-1\right)}
{8\left(4n+2l+1\right)\left(4n+2l+5\right)},\nonumber
\end{eqnarray}
\begin{equation}
\gamma_{n}^{ (3)}  =-
\frac{\text{b}^2(2n+2l+1)\
\left(n+l\right)}
{2\left(4n+2l-1\right)
\left(4n+2l+1\right)}.\hspace{1.5cm}
\end{equation}
and for $b_{n}^{(4)}$ the coefficients are
\begin{eqnarray}
\alpha_{n}^{ (4)} &=& -
 \frac{\text{b}^2(n+1)(2n+1)}
{2\left(4n+2l+3\right)\left(4n+2l+5\right)},\nonumber
\hspace{1.5cm}
\end{eqnarray}
\begin{eqnarray}
\beta_{n}^{ (4)} & = & {\cal E}-
\frac{\text{b}^2}{8}+\left( l-\frac{1}{2}\right)^2 +
4(n+l)\left(n+\frac{1}{2}\right)
\nonumber\\
&+&\frac{\text{b}^2\left(2l+1\right)
\left(2l-1\right)}
{8\left(4n+2l-1\right)\left(4n+2l+3\right)},\nonumber
\end{eqnarray}
\begin{eqnarray}
\gamma_{n}^{ (4)} & =&-
\frac{\text{b}^2(2n+2l-1)\
\left(n+l\right)}
{2\left(4n+2l-3\right)
\left(4n+2l-1\right)}.
\hspace{1.5cm}
\end{eqnarray}
According to section 2.3,  the corresponding expansions
given by series of the Bessel functions
$Y_{\alpha}$ and $H_{\alpha}^{(1,2)}$ converge
in the domain $|z|=\sinh^2u> 1$ which does not
include all the values of $u$.

On the other side, from the solutions $U_{7}^{(1)}$
and $U_{8}^{(1)}$ we find another pair of bounded
eigenfunctions given by
\antiletra
\begin{eqnarray}
\psi_{3}^{o}(u) &=&\tanh{u}
\displaystyle \sum_{n=0}^{\infty}
 \frac{\Gamma[n+(3/2)]}{(n+l)!}\ (-1)^n b_{n}^{(3)}
\nonumber\\
& \times& J_{2n+l+({3}/{2})}\left(\frac{\text{b}}{2}\cosh{u}\right),
\nonumber\end{eqnarray}
\begin{eqnarray}\label{infintas-3}
\psi_{3}^{\text{e}}(u) &=&
\displaystyle \sum_{n=0}^{\infty}
\frac{\Gamma[n+(1/2)]}{(n+l)!}\ (-1)^n b_{n}^{(4)}
\nonumber\\
&\times & J_{2n+l+({1}/{2})}\left(\frac{\text{b}}{2}\cosh{u}\right),
\end{eqnarray}
in which the series coefficients are proportional
to the coefficients of the previous pair and, consequently,
we have the same characteristic equations. Since
$\text{Re}(B_1/z_0)<1$ in equation (\ref{convergencia2}),
now the expansions in terms of
$Y_{\alpha}$ and $H_{\alpha}^{(1,2)}$ converge
in the domain $|1-z|=\cosh^2u\geq 1$ which
covers the entire range of $u$.

By using the solutions $U_{5}^{(j)}$ and $U_{6}^{(j)}$
we would find solutions equivalent the preceding ones.
Actually,  we would find
$\alpha_{n}^{ (5,6)}\propto (n-l+1)$  in the
recurrence relations for $b_{n}^{ (5)}$ and $b_{n}^{ (6)}$
what means that the series begin at $n=l$.
Thence, by setting $n=m+l$ we may conclude that
such solutions are proportional to the above ones.

At last, notice that the previous considerations
take into account only the analytical properties of the
wavefunctions. The full solution of the problem
requires the computation of the characteristic
equation resulting from the three-term recurrence
relations, which is represented by an infinite continued fraction
having the form given in equation (\ref{r1a})
or by the determinant of an infinite tridiagonal matrix.
%

%
%
\section{The electron
in the field of a point dipole}
For an electron with mass $M$, charge $\mathrm{e}$ and energy $E$
in the field of a point electric dipole,
the time-independent Schr\"odinger equation is \cite{levy}
 \begin{eqnarray*}
\left( -\frac{\hbar^2}{2M}\nabla^2+\mathrm{e}\frac{\vec{D}
\cdot \vec{r}}{r^3}-E\right)\psi=0,
 \end{eqnarray*}
where $D$ is the dipole moment. This equation is separable in
spherical coordinates ($r,\theta,\varphi$). Choosing the $z$
axis along the dipole moment and performing the separation
\begin{eqnarray*}
\Psi (r,\theta,\varphi)=\frac{1}{r} R(r)\ \Theta(\theta)\
e^{\pm im\varphi},\quad m=0,1,2,
\cdots
\end{eqnarray*}
($ 0\leq\theta\leq\pi$) one gets
\letra
\begin{equation}\label{dipolo}
\left[ \frac{1}{\sin{\theta}}\frac{d}{d\theta}\left(
\sin{\theta}\frac{d}{d\theta}\right)+C-\beta\cos{\theta}-
\frac{m^2}{\sin^2\theta}\right] \Theta=0,
\end{equation}
\begin{eqnarray}
\left[\frac{d^2 }{dr^2} - \frac{C}{r^2} + \cal E \right]R=0,
\end{eqnarray}
where $C$ is the separation constant,
$\beta=2M\mathrm{e}D/\hbar^2$ and ${\cal E}=
2ME/\hbar^2$. The energies are determined from the
solutions of the radial equation, but firstly it is necessary
to determine the parameter $C$ from the solutions
of the angular equation. The substitutions
\antiletra
\begin{eqnarray}\label{Teta}
\Theta(\theta)=(\sin{\theta})^{m}\ U(z),
\quad z=\sin^2\left({\theta}/{2} \right),
\end{eqnarray}
give the equation
\begin{eqnarray}\label{nonumber}\displaystyle
z(z-1)\frac{d^{2}U}{dz^{2}}-\big[m+1-2\left(m+1\right)z\big]
\frac{dU}{dz}
+\nonumber\\
\big[m\left(m+1\right)-\beta-C-2\beta(z-1)\big]U=0,
\end{eqnarray}
which is the Whittaker-Ince limit (\ref{incegswe}) of the CHE,
with the following set of parameters
%
\[z_{0}=1,\quad B_{1}=-m-1,\ \
B_2=2m+2,\]
\begin{equation}\label{parC0}
B_3=m\left(m+1\right)-\beta-C,\qquad  q=-2\beta.
\end{equation}
Therefore, the solutions $\Theta(\theta)$ can be
constructed by introducing solutions of equation
(\ref{incegswe}) into equation (\ref{Teta}). Then,
the admissible values
for $C$ are determined from the characteristic equations
which follow from the recurrence relations for the
series coefficients.

For $\beta=0$ the angular
equation (\ref{dipolo}) has solutions regular in
the interval $0\leq\theta\leq \pi$
if $C=\ell(\ell+1)$, where $\ell$ is a non-negative integer
such that $\ell\geq m$. These solutions are
given by the associated Legendre polynomials
$P_{\ell}^{m}(\cos\theta)$. For this trivial case,
a closed form for energy spectrum can be obtained
from boundary conditions on the radial
part of the wavefunction. However, if $\beta\neq 0$
there no analytic formula for $\cal E $
since the constant $C$ must be determined from a
transcendental equation (characteristic equation).

In section IV.A we write the expansions in series
of Bessel functions for the angular equation.
We find two periodic expansions having
period $2\pi$ and the same characteristic
equation. However, in section IV.B
we find only one expansion in series of associated
Legendre polynomial; this has the same characteristic
equation as the solutions in terms of Bessel functions.

\subsection{Expansions in series of Bessel functions}

For $\beta\neq 0$ the solutions $\Theta(\theta)$
can be obtained by inserting into (\ref{Teta}) the solutions
in series of Bessel functions of the first.

The other expansions are unsuitable in virtue
of their domain of convergence.

Thus, using the
parameters (\ref{parC0}) we find out that only
$U_{1}^{(1)}$, $U_{2}^{(1)}$,
$U_{5}^{(1)}$ and $U_{6}^{(1)}$ are valid.
Moreover, both $U_{1}^{(1)}$ and $U_{2}^{(1)}$
afford the same solution, denoted by  $\Theta_{1}$.
Analogously, both $U_5^{(1)}$ and $U_6^{(1)}$ yield
another solution, denoted by  $\Theta_{2}$. These solutions are
\begin{eqnarray}
\ \Theta_{1}(\theta)
&=&\left( \sin\frac{\theta}{2} \right)^{-1}
\left(  \cot\frac{\theta}{2}\right) ^{m}
\displaystyle \sum_{n=0}^{\infty}(-1)^n b_{n}^{(1)}\nonumber\\
&\times&J_{2n+2m+1} \left( i\sqrt{8\beta}\  \sin\frac{\theta}{2}
\right),
\end{eqnarray}
\begin{eqnarray}
\Theta_{2}(\theta)&=&
\left(  \cos\frac{\theta}{2}\right)  ^{-1}
\left( \tan\frac{\theta}{2} \right)  ^{m}
\displaystyle \sum_{n=0}^{\infty} b_{n}^{(1)}\times\ \ \nonumber\\
& &J_{2n+2m+1} \left( \sqrt{8\beta}\
\cos\frac{\theta}{2}\right),
\end{eqnarray}
with the recurrence relations for $b_{n}^{(1)}$ given by
($\alpha_{-1}=0$)
\begin{eqnarray}\label{rec-dipolo}
&&\frac{\beta(n+1)}{(2n+2m+3)}b_{n+1}^{(1)}-
\Big[n\big(n+2m+1\big)+\nonumber\\
&&m\big(m+1\big)-C\Big]b_{n}^{(1)}+
\frac{\beta(n+2m)}{(2n+2m-1)}b_{n-1}^{(1)}=0.\qquad
\end{eqnarray}
These solutions have the same coefficients $b_{n}^{(1)}$
because equation (\ref{pMathieu}) implies that $b_{n}^{(5)}=
(-1)^nb_{n}^{(1)}$. Thus, there is only one
characteristic equation to determine the values
of the parameter $C$.

The solutions $\Theta_{1}$ and $\Theta_{2 }$ are connected
by the substitutions $\theta\rightarrow\theta+\pi$ and
$\beta\rightarrow -\beta$ which leave invariant
the angular equation (notice that this is
equivalent to the change of
$b_n$ by $(-1)^nb_n$ in the
recurrence relations). Both are convergent
in the interval  $0\leq \theta\leq \pi$ and
are regular at the singular points $\theta=0$ and $\theta=\pi$.
We have found
no criterion to discard one of these solutions,
neither have found a proof that they are equivalent
(the problem mentioned in section 2.2).
However, in the following we will find only one
solution in series of Jacobi polynomials; this
has the same characteristic equation as the
above solutions.

\subsection{Expansions in series of Jacobi polynomials}
The expansions in series of Bessel functions given in
section II appear associated with a group of expansion
in series of Gauss hypergeometric
functions $F(a,b;c;x)$, which are obtained by
applying the transformations
rules to the solution \cite{eu}
\begin{equation}\label{legendre1}
\mathbb{U}_{1}(z)=
\displaystyle \sum_{n=0}^{\infty}b_{n}^{(1)}F\left(-n,
n+B_{2}-1;B_{2}+\frac{B_{1}}{z_{0}};
1-\frac{z}{z_{0}}\right),
\end{equation}
[$ B_{2}\neq 0,-1,-2,\cdots$] where $b_{n}^{(1)}$
is the same as in (\ref{bessel1b})
and satisfies the recurrence relations (\ref{bessel1d}).
This series converges for finite values of $z$.
The restrictions on the values of $B_2$ assure
independence of the hypergeometric functions.
However, we must demand as well that $c\neq$
$0,-1,-2,\cdots$ because in general
$F(a,b;c;x)$ is not defined if $c$ is
a negative integer or zero. For this reason, this
group is less general than the group formed by
series of Bessel functions. Now, by taking  $\mathbb{U}_{5}(z)=\mathscr{T}_{3}\mathbb{U}_{1}(z)$,
we find the solution
\begin{eqnarray}
\mathbb{U}_{5}(z)=
\displaystyle \sum_{n=0}^{\infty}b_{n}^{(5)}F\left(-n,
n+B_{2}-1;-\frac{B_{1}}{z_{0}};
\frac{z}{z_{0}}\right),\\
\left[B_{2}\neq 0,-1,-2,\cdots\right]\nonumber
\end{eqnarray}
where the coefficients $b_{n}^{(5)}$ are
formally connected with $b_{n}^{(1)}$
by equation (\ref{pMathieu}). If $B_2+B_{1}/z_0$
and ($-B_{1}/z_0$) are not zero or negative
integers, both solutions are valid and are given by
infinite series. Then, setting $z=z_0\cos^2(\sigma u)$,
using the relation  (\ref{pMathieu})
and rewriting the hypergeometric functions in terms of Jacobi's
polynomials $P_{n}^{(\alpha,\beta)}$
through equation (\ref{jacobi}), we find
\begin{eqnarray*}
\mathbb{U}_{1}(z)=\mathbb{W}_{1}(u)&=&
\Gamma(\alpha+1)
\displaystyle \sum_{n=0}^{\infty}\frac{n!\ b_{n}^{(1)}}
{\Gamma(n+\alpha+1)}\\
&\times& P_{n}^{(\alpha,\beta)}
\big[\cos(2\sigma u)\big],
\end{eqnarray*}
\begin{eqnarray*}
\mathbb{U}_{5}(z)=\mathbb{W}_{5}(u)&=&
\Gamma(\beta+1)
\displaystyle \sum_{n=0}^{\infty}\frac{(-1)^n\ n!\ b_{n}^{(1)}}
{\Gamma(n+\alpha+1)}\\
&\times& P_{n}^{(\beta,\alpha)}
\big[-\cos(2\sigma u)\big],
\end{eqnarray*}
where $\alpha=B_2-1+B_1/z_0$ and $\beta=-1-B_1/z_0$
(this $\beta$ should not be confused with the parameter
of the angular equation).
Then, relation (\ref{secao-4.2}) implies that
$\mathbb{U}_{1}$ is a multiple of $\mathbb{U}_{5}$.
This conclusion does not hold if
only one solution is valid. The same can be said of
the other pairs of solutions,
($\mathbb{U}_{i},\mathbb{U}_{i+4}$).
In this manner, the linear dependence of solutions
having the same characteristic equation is
almost trivial in the present case.

The solution $\mathbb{U}_{1}$
(equivalent to $\mathbb{U}_{5}$)
is the only one applicable to the angular equation of the
point dipole. Thus, by inserting this into (\ref{Teta})
and using the parameters (\ref{parC0}),
we find
\begin{eqnarray*}
\Theta(\theta)&=&\left(\sin\theta\right)^{m}
\displaystyle \sum_{n=0}^{\infty}b_{n}^{(1)}\\
&\times&F\big[-n,
n+2m+1;m+1;
\cos^2(\theta/2)\big],
\end{eqnarray*}
where the coefficients $b_{n}^{(1)}$ again satisfy the
recurrence relations (\ref{rec-dipolo}). In terms
of  Jacobi polynomials (\ref{jacobi})
or associated Legendre polynomials (\ref{legendre}),
we find that
\begin{eqnarray}\label{dipolo-1}
\Theta(\theta)&=&
\left(\sin\theta\right)^{m}
\displaystyle \sum_{n=0}^{\infty}\frac{(-1)^n  n!}{(n+m)!} b_{n}^{(1)}
P_{n}^{(m,m)}(\cos\theta)\nonumber\\
&\propto&
\displaystyle \sum_{n=0}^{\infty}\frac{(-1)^n n!}{(n+2m)!} b_{n}^{(1)}
P_{n+m}^{m}(\cos\theta).
\end{eqnarray}
If $m=0$, this reduces to the solutions
in series of ordinary Legendre
polynomials $P_n=P_{n}^{0}$
given by L\'evy-Leblond \cite{levy}. If $m\neq 0$,
these are the Alhaidari solutions
in terms of Jacobi polynomials \cite{alhaidari}
up to a redefinition of the series coefficients.
On the other side, by taking into account the
solutions $\mathbb{U}_{2}$,
$\mathbb{U}_{3}$ and $\mathbb{U}_{4}$,
written in \cite{eu}, we find that these are not
valid by one of the following reasons: the
last parameter of the hypergeometric functions
is a negative integer or $\Theta(\theta)$ is
not regular at $\theta=0$ or $\theta=\pi$.

Finally, the expansions in hypergeometric
functions are inapplicable to the Schr\"odinger equation
with the inverted potential (\ref{v1}).
In effect, infinite-series solutions are inappropriate because
they converge only for finite values of the arguments
of the hypergeometric
functions. On the other side, only $\mathbb{U}_{1}(z)$
and $\mathbb{U}_{2}(z)$
could afford finite-series solutions but in these cases
the hypergeometric functions are not defined since the
last parameter is zero or negative integer, that is,
$c=B_2+B_1/z_0=1-l$.
%
%
%
%
\section{Conclusion}

We have dealt with solutions for the Whittaker-Ince
limit (\ref{incegswe}) of the confluent Heun equation (CHE) and possible
applications of these solutions. Specifically,
in section II we have considered expansions in series
of Bessel functions
of the first and second kind, in addition to
solutions in series of Hankel functions
given in a previous paper \cite{eu}.
In sections III and IV we have established solutions in
series of Bessel functions for an inverted
potential and for an angular
equation for a point electric dipole, respectively.

Under certain conditions the expansions in terms
of Bessel functions of the first kind are convergent
and bounded for any value of $z$, including the point $z=\infty$.
These are suitable to solve angular equations
since in this case $z$ assumes finite values around
$z=0$. The other expansions in general converge
for $|z|> |z_0| $ or $
|z-z_0|> |z_0|$ but, under special conditions, may
converge also at $|z|=|z_0| $ or $
|z-z_0|=|z_0|$. For this reason, these and the solutions in
series of Bessel functions
of the first kind are applicable to the inverted potential.

We have noticed the possible coexistence
of different sets of expansions
in series of Bessel functions having the
same characteristic equation.
For the Mathieu equation it is known that these
solutions are linearly dependent in a common region
of convergence \cite{arscott,McLachlan}.
The same holds for expansions in series of Jacobi
polynomials, as we have shown in section IV.B.
If this dependence is valid for the general case,
we can avoid the duplicity of solutions in infinite
series of Bessel functions.
Nevertheless, we have
found no proof for such conjecture.

In section III we  have obtained eigenstates given by finite
and by infinite series of Bessel
functions for the one-dimensional Schr\"odinger
equation with the inverted quasi-exactly solvable
potential. For any value of the parameter $l$ which
characterizes the quasi-exact solvability, we have
proved that the degeneracy of
even and odd states given by finite series follows
from the three-term recurrence relations for the series coefficients.
These quasi-polynomial solutions permit to
determine only a part of the energy spectrum
by using, for instance, the procedure of Bender and Dunne
presented in section III.A.

We have found that the infinite-series solutions
are also convergent and bounded for any value of the
variable $u$. Odd and even solutions
are given by the expansions in series of Bessel functions
of the first kind, while other kinds of Bessel functions
give bounded solutions without definite parity.
In principle, the characteristic equations associated
with these solutions may afford
the remaining part of the energy spectrum.

In section IV we have found solutions given either
by series of Bessel functions or by series of
Jacobi  polynomials for the $\theta$-dependence
of the scattering of
electrons by the field of the point electric dipole.
The expansion in series of Jacobi polynomials
is equivalent to the one found by Alhaidari
\cite{alhaidari} and includes, as a particular case,
the solution in series of Legendre polynomials
proposed by L\'evi-Leblond for the case $m=0$  \cite{levy}.
We have found two expansions in series of Bessel
functions corresponding to the same characteristic equation.
We have also seen that the expansions
in series of Jacobi polynomials are inappropriate
for the inverted potential.


We observe that, by using transformations rules, we
can find different sets of solutions
with the same characteristic equation also for expansions
in series of confluent hypergeometric functions for
the CHE, given in \cite{eu0} -- these are the ones which lead to
the expansions in series of Bessel functions discussed in
section II. Then, the issue concerning
the linear dependence or independence of solutions arises
in this case as well.

Finally we remark that we have dealt only with one-sided
series solutions which are valid if the differential
equation has an arbitrary parameter. If there
is no free parameter, we have to deal with two-sided infinite
series solutions ($-\infty<n<\infty$).
For the CHE there are, for instance,
expansions in double-infinite series of hypergeometric
and confluent hypergeometric
functions \cite{leaver1,eu0} which admit of the
Whittaker-Ince limit \cite{eu}.
These solutions need to be extended
to incorporate all the Meixner expansions in series of Legendre and
Bessel functions  for the ordinary spheroidal wave
equation \cite{erdelyi1b,meixner}, that is, for the CHE
(\ref{gswe}) with $\eta=0$.

%
%
\appendix
\section{Some mathematical formulas}
\protect\label{A}
\setcounter{equation}{0}
\renewcommand{\theequation}{A.\arabic{equation}}

Firstly we give some properties of the Bessel functions
$J_{\alpha}(x)$, $Y_{\alpha}(x)$, $H_{\alpha}^{(1)}(x)$
and $H_{\alpha}^{(2)}(x)$.
After this we write some formulas concerning the
hypergeometric functions. The power-series representation
for $J_{\alpha}(x)$ is
\begin{eqnarray}\label{bessel}
J_{\alpha}(x)=
\left(\frac{x}{2} \right)^{\alpha} \sum_{k=0}^{\infty}
\frac{(-1)^k}{k! \ \Gamma(\alpha+k+1)}{\left({\frac{x}{2}}\right)}^{2k}.
\end{eqnarray}
This function is connected with the Hankel functions by the relation
\begin{eqnarray}\label{J-H}
J_{\alpha}(x)=\frac{1}{2}\big[H_{\alpha}^{(1)}(x)+
H_{\alpha}^{(2)}(x)\big].
\end{eqnarray}
Similarly, the Bessel functions of the second kind
can be expressed as
\begin{eqnarray}\label{Y-H}
Y_{\alpha}(x)=\frac{1}{2i}\big[H_{\alpha}^{(1)}(x)-
H_{\alpha}^{(2)}(x)\big],
\end{eqnarray}
On the other hand, for a fixed $\alpha$ the
behaviors of the Bessel functions when
$|x|\rightarrow\infty$ are given by  \cite{erdelyi1b}
\begin{equation}
J_{\alpha}(x)\sim
\sqrt{\frac{2}{\pi x}}
  \cos\left( x-\frac{1}{2}\alpha\pi-\frac{1}{4}\pi\right),
\ |\arg\ x|<\pi;\nonumber\qquad
\end{equation}
\begin{equation}
H_{\alpha}^{(1)}(x)\sim\sqrt{\frac{2}{\pi x}}
e^{ {i}( x-\frac{1}{2}\alpha\pi-\frac{1}{4}\pi)},
\ -\pi<\arg\ x<2\pi;\nonumber\qquad\
\end{equation}
\begin{equation}\label{Bessel-assimptotico}
H_{\alpha}^{(2)}(x)\sim\sqrt{\frac{2}{\pi x}}
e^{ -{i}( x-\frac{1}{2}\alpha\pi-\frac{1}{4}\pi)},
\  -2\pi<\arg\ x<\pi.\qquad
\end{equation}
The behavior for $Y_{\alpha}(x)$ is obtained by
changing the cosines by sines in the expression for
$J_{\alpha}(x)$. The Bessel functions are given by
finite series of elementary functions if their order
$\alpha$ is half an odd integer; then the restrictions
on $\arg{x}$ are unnecessary in (\ref{Bessel-assimptotico})
\cite{luke}.  For instance,
if $m=0,1,2,\cdots$, the Bessel functions of the first kind
can be expressed as
\begin{equation}\label{esferica1}
J_{-m-\frac{1}{2}}(x)=
\sqrt{\frac{2}{\pi }}\ x^{m+\frac{1}{2}}\left(\frac{d}{xdx} \right)^{m}
\frac{\cos{x}}{x},\nonumber\hspace{1.4cm}
\end{equation}
\begin{equation}
J_{m+\frac{1}{2}}(x)=\sqrt{\frac{2}{\pi }}\ (-1)^{m}
 x^{m+\frac{1}{2}}\left(\frac{d}{xdx} \right)^{m}
\frac{\sin{x}}{x}
\end{equation}
which give, in particular,
\begin{eqnarray}\label{bessel-1/2}
&& J_{-\frac{1}{2}}\left(x\right)=\sqrt{\frac{2}{\pi x}}\ \cos x,\ \
  J_{\frac{1}{2}}\left(x\right)=\sqrt{\frac{2}{\pi x}}\ \sin x,
\nonumber
\end{eqnarray}
\begin{eqnarray}
&& J_{-\frac{3}{2}}\left(x\right)=
\sqrt{\frac{2}{\pi x}}\left(-\sin{x}-\frac{\cos{x}}{x}\right),
\nonumber\hspace{1.7cm}
\end{eqnarray}
\begin{eqnarray}
&&J_{\frac{3}{2}}\left(x\right)=
  \sqrt{\frac{2}{\pi x}}\left(\frac{\sin{x}}{x}-\cos{x}\right).
\hspace{2.2cm}
\end{eqnarray}
Since, for $x$ fixed and $\alpha\rightarrow \infty$,
\begin{eqnarray*}
J_{\alpha}(x)\sim\frac{1}{\Gamma(\alpha+1)}
\left( \frac{x}{2}\right)^{\alpha},\hspace{3cm}
\end{eqnarray*}
\begin{eqnarray*}
&&Y_{\alpha}(x)\sim -iH_{\alpha}^{(1)}(x)
\sim iH_{\alpha}^{(2)}(x)\sim-\frac{1}{\pi}\Gamma(\alpha)
\left( \frac{2}{x}\right)^{\alpha},
\end{eqnarray*}
we find the relations
\begin{eqnarray}\label{JY}
\frac{J_{\alpha+2}(x)}{J_{\alpha}(x)}\sim
\frac{x^2}{4(\alpha+1)(\alpha+2)},\hspace{1.2cm}
\nonumber
\end{eqnarray}
\begin{eqnarray}
\frac{Z_{\alpha+2}^{(j)}(x)}{Z_{\alpha}^{(j)}(x)}\sim
\frac{4\alpha(\alpha+1)}{x^2},\quad(j=2,3,4)
\end{eqnarray}
which have been used in section II.C.

 In section IV, the relation between
hypergeometric functions and the Jacobi polynomials
$P_{n}^{(\alpha,\beta)}$ was obtained from \cite{abramowitz}
\begin{eqnarray}\label{jacobi}
F(-n,n+1+\alpha+\beta;1+\alpha;y)=\nonumber\\
\frac{n!\
\Gamma(\alpha+1)}{\Gamma(n+\alpha+1)}\
P_{n}^{(\alpha,\beta)}(1-2y),
\end{eqnarray}
where $n$ is a non-negative integer.
The Jacobi polynomials can be expressed as
\begin{eqnarray}
 P_n^{(\alpha,\beta)}(x)&=&\frac{(-1)^n}{2^nn!}(1-x)^{-\alpha}
(1+x)^{-\beta}\nonumber\\
&\times&\frac{d^n}{dx^n}\left[(1-x)^{\alpha+n}
(1+x)^{\beta+n}\right],
 \end{eqnarray}
whereby we find the relation
\begin{eqnarray}\label{secao-4.2}
 P_n^{(\beta,\alpha)}(-x)=(-1)^n P_n^{(\alpha,\beta)}(x),
\end{eqnarray}
used in section 4.2.

On the other side, the relation between
hypergeometric functions and the associated Legendre functions
$P_{\nu}^{k}$, when $k$ is a positive integer or zero, is given by
\begin{eqnarray}\label{legendre}
F(k-\nu,\nu+k+1;k+1;y)=
\frac{(-1)^k k!
\Gamma(\nu+1-k)}{\Gamma(\nu+k+1)}\nonumber\\
\times \left(y-y^2\right)^{-\frac{k}{2}}
P_{\nu}^{k}(1-2y).\qquad
\end{eqnarray}
Thence, by setting $\alpha=\beta=k$ in (\ref{jacobi}) and $\nu=n+k$
in (\ref{legendre}) we obtain
\begin{eqnarray}\label{jacobi-legendre}
P_{n}^{(k,k)}(\xi)=\frac{(-2)^k(n+k)!}{(n+2k)!}%
\left(1-\xi^2\right)^{-k/2}P_{n+k}^{k}(\xi).\nonumber\\
\end{eqnarray}
where we have put $\xi=1-2y$.
%

\section{The other solutions for equation (\ref{incegswe})}
\protect\label{B}
\setcounter{equation}{0}
\renewcommand{\theequation}{B.\arabic{equation}}
%
%
%
From the solutions (\ref{bessel1b}), the others are
obtained by using the transformations
$\mathscr{T}_1$ and $\mathscr{T}_2$
as indicated in the sequences (\ref{sequence2})
and (\ref{sequence3}). In this manner, we get
\begin{eqnarray}
U_{2}^{(j)}(z)&=&z^{(1-B_{2})/{2}}
\displaystyle \sum_{n=0}^{\infty}(-1)^nb_{n}^{(2)}\nonumber\\
& \times&Z_{2n+1+B_{2}+({2B_{1}}/{z_{0}})}^{(j)}\left(2\sqrt{qz}
\right),\nonumber\hspace{1.7cm}
\end{eqnarray}
\begin{eqnarray}\label{bessel2b}
U_{6}^{(j)}(z) &=&z^{1+\frac{B_1}{z_0}}
\big(z-z_{0}\big)^{-\frac{1}{2}-\frac{B_1}{z_0}-\frac{B_2}{2}}
\displaystyle \sum_{n=0}^{\infty} (-1)^nb_{n}^{(6)}
\nonumber\\
&\times&Z_{2n+1+B_{2}+({2B_1}/{z_0})}^{(j)}\left(2\sqrt{q(z-z_0)}\right),
\\
& & \hspace{1cm}  \left[\frac{B_{2}}{2}+\frac{B_{1}}{z_{0}}\neq-1,
-\frac{3}{2},-2,\cdots\right]\nonumber
\end{eqnarray}
where
\begin{eqnarray}
\alpha_{n}^{^{(2)}}=  \frac{qz_{0} (n+1)
\left(n+2+\frac{B_{1}}{z_{0}}\right)}
{\left(n+1+\frac{B_{2}}{2}+\frac{B_{1}}{z_{0}}\right)
\left(n+\frac{3}{2}+\frac{B_{2}}{2}+\frac{B_{1}}{z_{0}}
\right)},\nonumber\
\end{eqnarray}
\begin{eqnarray}
&&\beta_{n}^{^{(2)}}=
4B_{3}-2q z_{0}+4\left[n+1+\frac{B_{1}}{z_{0}}\right]
\left[n+B_{2}+\frac{B_{1}}{z_{0}}\right]\
\nonumber\\
&&\qquad-\frac{2q z_{0}\left(\frac{B_{2}}{2}-1\right)
\left(\frac{B_{2}}{2}+\frac{B_{1}}{z_{0}}\right)}
{\left(n+\frac{B_{2}}{2}+\frac{B_{1}}{z_{0}}\right)
\left(n+1+\frac{B_{2}}{2}+\frac{B_{1}}{z_{0}}\right)},
\nonumber
\end{eqnarray}
\begin{equation}\label{bessel2c}
\gamma_{n}^{^{(2)}}= \frac{q z_{0}\left[n+B_{2}+
\frac{B_{1}}{z_{0}}-1\right]
\left[n+B_{2}+\frac{2B_{1}}{z_{0}}\right]}
{\left(n-\frac{1}{2}+\frac{B_{2}}{2}+\frac{B_{1}}
{z_{0}}\right)
\left(n+\frac{B_{2}}{2}+\frac{B_{1}}{z_{0}}\right)},
\hspace{6.5mm}
\end{equation}
in the recurrence relations for $b_{n}^{(2)}$, which
are given by the equations
\begin{eqnarray}\label{bessel2d}
&&\text{Eq. (\ref{r1a})
if }\frac{B_{2}}{2}+\frac{B_{1}}{z_{0}}\neq 0,
-\frac{1}{2}; \  \nonumber
\end{eqnarray}
\begin{eqnarray}
&&\text{(\ref{r2a}) if }
\frac{B_{2}}{2}+\frac{B_{1}}{z_{0}}=
-\frac{1}{2};\
  \text{(\ref{r3a}) if }
\frac{B_{2}}{2}+\frac{B_{1}}{z_{0}}=0.\qquad
\end{eqnarray}
%
These relations also hold for
the coefficients $b_{n}^{(6)}$ with $\beta_{n}^{(6)}  = \beta_{n}^{(2)}$ and
\begin{eqnarray}
\alpha_{n}^{(6)}  =  \frac{-q z_{0}\left(n+1\right)
\left(n+B_2+\frac{B_{1}}{z_{0}}\right)}
{\left(n+\frac{B_1}{z_0}+\frac{B_{2}}{2}+\frac{1}{2}\right)
\left(n+\frac{B_1}{z_0}+\frac{B_{2}}{2}+\frac{3}{2}
\right)} ,\nonumber
\end{eqnarray}
\begin{eqnarray}
\gamma_{n}^{(6)} =- \frac{ q z_{0}\left(n+B_{2}+\frac{2B_1}{z_0}\right)
\left(n+1+\frac{B_{1}}{z_{0}}\right)}
{\left (n-\frac{1}{2}+\frac{B_1}{z_0}+\frac{B_{2}}{2}\right)
\left(n+\frac{B_1}{z_0}+\frac{B_{2}}{2}\right)},\quad
%
\end{eqnarray}
in the recurrence relations  (\ref{bessel2d}) for
the coefficients $b_{n}^{(6)}$.

%
%
%
For the third and seventh sets we find
\begin{eqnarray}
U_{3}^{(j)}(z) &=&(z-z_{0})^{1-B_{2}-
\frac{B_{1}}{z_{0}}}\
z^{\frac{B_{1}}{z_{0}}+\frac{B_{2}}{2}-\frac{1}{2}}
\nonumber\\
&\times&
\displaystyle \sum_{n=0}^{\infty}(-1)^{n}b_{n}^{(3)}
Z_{2n+3-B_{2}}^{(j)}\left(2\sqrt{qz}\right),\nonumber
\hspace{13mm}
\end{eqnarray}
\begin{eqnarray}
U_{7}^{(j)}(z) &=&z^{1+
\frac{B_{1}}{z_{0}}}\
\big(z-z_0\big)^{-\frac{1}{2}-\frac{B_{1}}{z_{0}}-\frac{B_{2}}{2}}
\displaystyle \sum_{n=0}^{\infty}(-1)^{n}b_{n}^{(7)}
\nonumber\\
&\times&
Z_{2n+3-B_{2}}^{(j)}\left(2\sqrt{q(z-z_0)}\right),\\
&  &\hspace{3cm}\left[B_{2}\neq 4,5,6,\cdots \right]
\nonumber
\end{eqnarray}
with the coefficients
\begin{eqnarray}
\alpha_{n}^{ (3)}& =&  \frac{qz_{0}\
(n+1)
\left(n+2+\frac{B_{1}}{z_{0}}\right)}
{\left(n+2-\frac{B_{2}}{2}\right)\left(n+\frac{5}{2}-
\frac{B_{2}}{2}\right)},\nonumber
\hspace{2cm}
\end{eqnarray}
\begin{eqnarray}
\beta_{n}^{ (3)} & = & 4B_{3}-2q z_{0}+4(n+1)(n+2-B_{2})\hspace{1cm}\nonumber\\
& -&\frac{2q z_{0}\left(\frac{B_{2}}{2}-1\right)
\left(\frac{B_{2}}{2}+\frac{B_{1}}{z_{0}}\right)}
{\left(n+1-\frac{B_{2}}{2}\right)\left(n+2-
\frac{B_{2}}{2}\right)},\nonumber
\end{eqnarray}
\begin{eqnarray}
\gamma_{n}^{ (3)} & =&
\frac{q z_{0}\ \left(n+2-B_{2}\right)
\left(n+1-B_{2}-\frac{B_{1}}{z_{0}}\right)}
{\left(n+\frac{1}{2}-\frac{B_{2}}{2}\right)
\left(n+1-\frac{B_{2}}{2}\right)}.\quad
\end{eqnarray}
in the recurrence relations for $b_{n}^{(3)}$ which are given by
\begin{eqnarray}\label{bessel3d}
&&\text{Eq. (\ref{r1a}) if }
B_{2}\neq 2,3;\ \text{ Eq. (\ref{r2a}) if } B_{2}=3;\nonumber
\end{eqnarray}
\begin{eqnarray}
&& \text{Eq. (\ref{r3a}) if } B_{2}=2.
\end{eqnarray}
%
%
%
We obtain the coefficients $\beta_{n}^{ (7)}  = \beta_{n}^{ (3)}$
and
\begin{equation}
\alpha_{n}^{ (7)} = - \frac{qz_{0}\
(n+1)
\left(n+2-B_2-\frac{B_{1}}{z_{0}}\right)}
{\left(n+2-\frac{B_{2}}{2}\right)\left(n+\frac{5}{2}-
\frac{B_{2}}{2}\right)},\nonumber\hspace{9mm}
\end{equation}
\begin{equation}
\gamma_{n}^{ (7)}  =-
\frac{q z_{0}\ \left(n+2-B_{2}\right)
\left(n+1+\frac{B_{1}}{z_{0}}\right)}
{\left(n+\frac{1}{2}-\frac{B_{2}}{2}\right)
\left(n+1-\frac{B_{2}}{2}\right)}.
\end{equation}
in the recurrence relations (\ref{bessel3d}) for
$b_{n}^{(7)}$.
%
%
%
%
Finally,
\begin{eqnarray}
U_{4}^{(j)}(z)&=&(z-z_{0})^{1-B_{2}-\frac{B_{1}}{z_{0}}}
z^{\frac{B_{1}}{z_{0}}+\frac{B_{2}}{2}-\frac{1}{2}}
\nonumber\\
& \times&\displaystyle \sum_{n=0}^{\infty}(-1)^{n}b_{n}^{ (4)}Z_{2n+1-B_{2}-({2B_{1}}/{z_{0}})}^{(j)}\left(2\sqrt{qz}\right),
\nonumber
\end{eqnarray}
\begin{eqnarray}
U_{8}^{(j)}(z)&=&z^{(1-B_{2})/2}
\displaystyle \sum_{n=0}^{\infty}(-1)^{n}b_{n}^{ (8)}
\times\nonumber\\
& &Z_{2n+1-B_{2}-({2B_{1}}/{z_{0}})}^{(j)}
\left(2\sqrt{q(z-z_0)}\right),\qquad
\\
&  &  \hspace{.5cm}\left[\frac{B_{2}}{2}+\frac{B_{1}}{z_{0}}\neq1,\frac{3}{2},
2,\frac{5}{2}, \cdots\right]\nonumber
\end{eqnarray}
with
\begin{eqnarray}
\alpha_{n}^{ (4)}  =  \frac{qz_{0}\ (n+1)
\left(n-\frac{B_{1}}{z_{0}}\right)}
{\left(n+1-\frac{B_{2}}{2}-\frac{B_{1}}{z_{0}}\right)
\left(n+\frac{3}{2}-\frac{B_{2}}{2}-\frac{B_{1}}{z_{0}}
\right)},\nonumber
\end{eqnarray}
\begin{eqnarray}
&&\beta_{n}^{ (4)}  =
 4B_{3}-2q z_{0}+
4\left[n-\frac{B_{1}}{z_{0}}\right]\left[n-B_{2}+1-\frac{B_{1}}{z_{0}}\right]
\nonumber\\
&&\qquad-\frac{2q z_{0}\left(\frac{B_{2}}{2}-1\right)
\left(\frac{B_{2}}{2}+\frac{B_{1}}{z_{0}}\right)}
{\left(n-\frac{B_{2}}{2}-\frac{B_{1}}{z_{0}}\right)
\left(n+1-\frac{B_{2}}{2}-\frac{B_{1}}{z_{0}}\right)},
\nonumber
\end{eqnarray}
\begin{eqnarray}
\gamma_{n}^{ (4)}  = \frac{q z_{0}
\left[n+1-B_{2}-\frac{B_{1}}{z_{0}}\right]
\left[n-B_{2}-\frac{2B_{1}}{z_{0}}\right]}
{\left(n-\frac{1}{2}-\frac{B_{2}}{2}-\frac{B_{1}}
{z_{0}}\right)
\left(n-\frac{B_{2}}{2}-\frac{B_{1}}{z_{0}}\right)},
\quad
\end{eqnarray}
in the following recurrence relations
\begin{eqnarray}
&&\text{Eq. (\ref{r1a}) if }
\frac{B_{2}}{2}+\frac{B_{1}}{z_{0}}\neq 0, \frac{1}{2};\
\text{(\ref{r2a}) if }\frac{B_{2}}{2}+\frac{B_{1}}{z_{0}}=\frac{1}{2};\nonumber
\end{eqnarray}
\begin{eqnarray}\label{bessel4d}
&&\text{ (\ref{r3a}) if } \frac{B_{2}}{2}+\frac{B_{1}}{z_{0}}=0.
\end{eqnarray}
%
The coefficients $b_{n}^{(8)}$ also satisfy the recurrence
relations  (\ref{bessel4d}) with $ \beta_{n}^{ (8)}
=\beta_{n}^{ (4)}$ and
\begin{eqnarray}
\alpha_{n}^{ (8)}  =  \frac{-qz_{0}\ (n+1)
\left(n+2-B_2-\frac{B_{1}}{z_{0}}\right)}
{\left(n+1-\frac{B_{2}}{2}-\frac{B_{1}}{z_{0}}\right)
\left(n+\frac{3}{2}-\frac{B_{2}}{2}-\frac{B_{1}}{z_{0}}
\right)},\nonumber
\end{eqnarray}
\begin{eqnarray}
\gamma_{n}^{ (8)}  = -\frac{q z_{0}\
\left(n-1-\frac{B_{1}}{z_{0}}\right)
\left(n-B_{2}-\frac{2B_{1}}{z_{0}}\right)}
{\left(n-\frac{1}{2}-\frac{B_{2}}{2}-\frac{B_{1}}
{z_{0}}\right)
\left(n-\frac{B_{2}}{2}-\frac{B_{1}}{z_{0}}\right)}.
\end{eqnarray}
\vspace{1.0cm}
%
%

%

\end{document}